\documentclass[]{pasj02} 
\usepackage[switch,mathlines]{lineno} 
\usepackage{natbib} 

\jyear{2024}
\Received{}
\Accepted{}

\usepackage[OT2, T1]{fontenc}

\begin{document}

\newcommand{\vdag}{(v)^\dagger}
\newcommand\aastex{AAS\TeX}
\newcommand\latex{La\TeX}
\newcommand{\blue}[1]{\textcolor{blue}{\textbf{#1}}}
\newcommand{\magenta}[1]{\textcolor{magenta}{\textbf{#1}}}
\newcommand{\red}[1]{\textcolor{red}{\textbf{#1}}}
\newcommand\ergs{erg~s$^{-1}$}
\newcommand\sun{\odot}

\title{3D Moving-mesh Hydrodynamical Simulations of Wind/Jet Driven Ultraluminous X-ray Source Bubbles}

\author{Jiahui \textsc{Huang} \altaffilmark{1}}
\altaffiltext{1}{Center for Computational Sciences, University of Tsukuba, 1-1-1 Ten-nodai, Tsukuba, Ibaraki 305-8577, Japan}
\email{huangjh96@ccs.tsukuba.ac.jp}

\author{Ken \textsc{Ohsuga}\altaffilmark{1}}
\email{ohsuga@ccs.tsukuba.ac.jp}

\author{Hua \textsc{Feng}\altaffilmark{2}}
\altaffiltext{2}{State Key Laboratory of Particle Astrophysics, Institute of High Energy Physics, Chinese Academy of Sciences, Beijing 100049, China}
\email{hfeng@ihep.ac.cn}

\author{Hui \textsc{Li}\altaffilmark{3}}
\altaffiltext{3}{Department of Astronomy, Tsinghua University, Beijing 100084, China}
\email{hliastro@tsinghua.edu.cn}

\maketitle

\begin{abstract}

We perform 3 dimensional moving-mesh hydrodynamical simulations of bubble nebulae around ultraluminous X-ray sources, using state-of-the-art software AREPO. We use a Monte-Carlo method to inject outflows with uniform mass outflow rate and momentum, in a conical funnel with a specific half opening angle. Simulation results show that the morphology of the bubble is determined by the initial momentum of the outflows, while the mechanical power of the outflows only influences the size of the bubble without changing its shape. Low mechanical power also results in a short cooling timescale of the system, leading to an early collapse of the bubble shell. The half opening angle of the outflows and the viewing angle of the system determine the observed bubble eccentricity together. Compared with the observational morphology of the ULX bubble sources NGC~55~ULX-1 and NGC~1313~X-2, our simulation favors the fact that the high velocity outflows of the accretion disks in these two systems are confined in a narrow funnel region.

\end{abstract}

\section{Introduction} \label{sec:intro}

Ultraluminous X-ray sources (ULXs) are off-nuclear objects in external galaxies. They show an X-ray luminosity that exceeds $10^{39}$~\ergs, which corresponds to the Eddington luminosity of a 10~$M_{\sun}$ black hole \citep{Kaaret2017}. Discoveries of the pulsations in some ULXs demonstrate that these ULXs are powered by accretion onto neutron stars \citep{Bachetti2014, Furst2016, Israel2017, Israel2017a, Carpano2018, Sathyaprakash2019, Rodriguez2020}. Thus, these systems are experiencing super-Eddington accretion. There is also the possibility that the central object of a ULX is a BH; however, even in that case, if it is a stellar-mass BH, it must be accreting at super-Eddington rates (with one reported case suggesting an intermediate-mass black hole \citep{Sutton2012}).

During the super-Eddington accretion, it is theoretically expected that the strong radiation pressure will launch disk winds \citep{Shakura1973, Meier1982, Poutanen2007}. Observationally, some blue shifted absorption features in the ULX spectra demonstrate the high velocity outflows with $\sim0.1c$ from the accreting system \citep{Walton2016, Pinto2016, Kosec2018, Kosec2018a, Pinto2021}. In addition, the numerical simulations also reveal some outflow structures from the accretion disks. For example, 3-dimensional radiation magneto-hydrodynamic (3D-RMHD) simulations show strong radiation driven outflows with a velocity of 0.1--0.4$c$ \citep{Jiang2014, Sadowski2016, Takahashi2016, Huang2023, Asahina2024}, while 2-dimensional radiation hydrodynamic (2D-RHD) simulations show how the outflows eventually develop to a large radius of $\sim1000$ Schwarzschild radius \citep{Ohsuga2011, Hashizume2015, Kitaki2018, Kitaki2021, Yoshioka2022, Yoshioka2024}. Specifically, simulations of stellar mass black hole accretion, corresponding to the case of ULXs, show that the outflows can be either disk winds at large disk radii with a large half opening angle \citep{Ohsuga2005, Kitaki2021, Yoshioka2022}, or narrow angle collimated jet-like outflows with a small half opening angle around a spinning black hole \citep{Ohsuga2011, Sadowski2016, Asahina2024}, or a combination of high velocity collimated jet and wide-angle moderate velocity winds \citep{Jiang2014}. The simulations also predict that the strong outflows can carry a mechanical power at the level of $\sim10\%$ of the X-ray radiative luminosity when the accretion rate is supercritical \citep{Yoshioka2022}, which is consistent with the observations of the ULXs IC~342~X-1 \citep{Cseh2012} and Holmberg~II~X-1 \citep{Cseh2014, Cseh2015}.

The disk winds, which carry a large amount of mass and momentum, can interact with the interstellar medium (ISM) and create shock-ionized bubble nebulae. Such nebulae have been found in the optical bands around several ULXs \citep{Pakull2002, Pakull2003, Ramsey2006, Abolmasov2007, Russell2011, Soria2021, Zhou2022, Zhou2023, Guo2023} with a size around $\sim100$~pc and an expanding velocity of $\sim100$~km~s$^{-1}$. Combining the bubble size, the optical luminosity and the expanding velocity, the total mechanical power that drives the bubbles is estimated to be $10^{39}$--$10^{40}$~\ergs \citep{Weaver1977}. However, although we can infer the energetics of the ULX bubbles by applying the simplified self similar solutions, there are still mounting observations and details, especially those obtained with the integral field units (IFU) observations, need to be explained. Through detailed analysis of the IFU spectral line data, we can potentially infer the outflow structures from the embedded accretion systems from the line-of-sight velocity map and the velocity dispersion map \citep{Zhou2022, Zhou2023}.

In order to explain the detailed structures of the IFU observations, we can use the numerical simulations to explore a large parameter space of the underlying outflow properties and the ISM properties, and compare the resulting bubble structures with the IFU observations. By analyzing the simulated bubble structures, we can have a better understanding of the mechanisms driving these ULX bubbles, which can serve as a guide for the future IFU observations upon the ULX bubble nebulae. In addition, the simulations of the ULX bubble propagation can be easily related to the previous accretion disk simulations, by adopting the simulated angular distribution of the mass outflow rate and the momentum outflow rate as the input boundary conditions of the bubble formation. Therefore, by comparing the different bubble structures of the wide-angle outflows and the narrow funnel outflows with the observed bubble nebulae, we may further constrain the parameter space of the accretion disk simulations in return, which is difficult to achieve using the accretion disk scale simulations alone.

In previous literature, simulations of the large-scale effects due to the accretion outflows concentrate mostly on their feedback effects towards the galaxy evolution, e.g. AGN outflow feedback \citep{Springel2005, Booth2009, Weinberger2017a} and high-mass X-ray binary outflow feedback \citep{Artale2015}. Simulations of dynamical propagation of the outflows are mostly limited to the $\sim$kpc scale cavities produced by collimated jets from super-massive black holes \citep{Weinberger2017, Ohmura2023}, galaxy superbubbles \citep{Mou2014, Costa2020}, or jet-driven bubbles around some famous Galactic microquasars such as SS~433 \citep{Goodall2011, Monceau-Baroux2015}. By now, few studies, except for one 1-D simulation work \citep{Siwek2017}, have specifically focused on the simulations of ULX bubbles, which are also important sources showing the large-scale propagation of accretion outflows. Therefore, it is necessary to carry out pilot 3 dimensional simulations to systematically study the large-scale propagation of different types of ULX outflows (wide angle or collimated). In addition, \citet{Kitaki2021} have shown that the ratio of kinetic and radiative luminosities $L_{\rm kin}/L_{\rm rad}$ can be used to estimate the mass accretion rate of some specific ULXs. However, this ratio could change over large-scale propagation of the outflows. Performing large-scale simulations of outflow propagation can also help to clarify this point as well.

Different ULX bubbles possess a variety of morphologies. We expect that the bulk flows, turbulence, and density inhomogeneity of the surrounding ISM will have influence on the expansion and morphology of the bubble nebulae. Most bubbles show regular elliptical shapes, such as the nebulae around NGC~1313~X-2 \citep{Pakull2006, Zhou2022}, NGC~55~ULX-1 \citep{Zhou2023}, Ho~IX~X-1 \citep{Ramsey2006}, and S26 in NGC~7793 \citep{Pakull2010, Soria2010}. However, a part of the ULX bubbles have irregular morphologies, e.g. tooth shape nebula around IC~342~X-1 \citep{Cseh2012}, and foot shape nebula around Ho~II~X-1 \citep{Pakull2002}, which reflects that the central ULXs are embedded in a complex ISM environment. In this pilot study, for simplicity, we choose to focus on the regular shaped ULX bubbles, because it is probable that these bubbles reside in a moderately homogeneous ISM environment and contain relatively simple production mechanisms. We aim at using numerical simulations to explore the dependencies of different outflow parameters on the properties of ULX bubbles. 

We organize this paper as follows. In Section~\ref{sec:methods}, we describe the setups of our simulation framework and the outflow injection method. In Section~\ref{sec:result}, we display the results of our simulation and discuss about the parameter dependencies. In Section~\ref{sec:discussion}, we discuss about how our results can shed light on the current observational data. And finally in Section~\ref{sec:conclusion}, we summarize the conclusions of this work.

\section{Methods} \label{sec:methods}

\subsection{Simulation setup} \label{sec:setup}

We perform our simulation using the moving-mesh, finite-volume hydrodynamic software AREPO \citep{Springel2010, Weinberger2020}. The simulation domain is a $400\times400\times400$~pc box, with $2\times256^3$ mesh-generating points inside. We test the convergence with spatial resolutions by performing our fiducial run using $2\times128^3$ and $2\times64^3$ mesh-generating points. We find that the eccentricity of the bubble and the thickness of the bubble shell are consistent in the two low resolution runs with the fiducial run. Thus we conclude that the results are converged with $2\times256^3$ mesh-generating points. These mesh-generating points will form a Voronoi tessellation from the dual Delaunay tessellation method. During the simulation, the mesh-generating points will follow the gas flow and move within the simulation domain in a quasi-Lagrangian way. We will (de)refine a gas cell if the mass of the cell is larger (smaller) than twice (half) of the target mass defined by the initial conditions, to keep the mass of all the gas cells close to the target mass. The target mass is defined as $M_{\rm target} = \frac{\mu m_{\rm p} n_{\rm ISM} V_{\rm box}}{N_{\rm cell}}$, where $n_{\rm ISM}$ is the initial ISM number density, $V_{\rm box}=400\times400\times400$~pc$^3$ is the simulation box volume, and $N_{\rm cell}=2\times256^3$ is the initial cell number. In addition, we will refine a cell when the volume of the cell is larger than 32 times of all its adjacent cells, to prevent large volume contrast \citep{Li2019}. The refinement of the gas cells will ensure that a better resolved meshgrid will be generated at the dense shock front produced by the accretion disk winds, and trace its propagation automatically.

We include basic hydrodynamics and self-gravity in our simulation, and use an adaptive softening scheme to resolve the gravitational forces down to each gas cell. We perform our simulation runs using the tree-based timestep method implemented in AREPO. We perform regularization of the mesh at regular intervals to keep the cells relatively round. Due to our wind injection procedure, we find that elongated cells appear near the edge of our injection spheres. If one cell satisfies that the distance between center of mass and mesh-generating point exceeds the effective radius, we define it as highly-distorted, where the effective radius is defined as the radius of a sphere having the same volume as the cell. However, these are very few number (0.02\% of all cells), and thus unlikely to introduce large errors. We handle the radiative cooling in an explicit way from multiple channels, including the heating and cooling processes of hydrogen and helium due to recombinations, collisions, and free-free emission; high-temperature ($T>10^4$~K) metal-line cooling following the recipe of \citet{Vogelsberger2013}; low-temperature ($T<10^4$~K) metal-line, fine-structure, and molecular cooling following the fitted function of temperature, density, and gas metallicity calculated from the table from CLOUDY \citep{Hopkins2018, Marinacci2019}. We set a minimum gas temperature of 100~K so that the gas will only be allowed to cool down to this temperature radiatively. In all of our simulation runs, we find that the minimum gas temperature is at a level of $\sim$1000~K, far from the floor value. We adopt the periodic boundary conditions at each outer surface of the simulation box.

In this work, we neglect the effect of X-ray radiative force, because it is much smaller than the ram pressure force. Under the assumption that the system is spherically symmetric and optically thin, the radiative force can be approximated as $f_{\rm rad}=\frac{\chi L_{\rm X}}{4\pi r^2 c}$, where $\chi$ is the total opacity \citep{Ohsuga2011}. By adopting a typical central X-ray luminosity of $L_{\rm X} \sim 10^{39}$~\ergs, we find that $f_{\rm rad}/f_{\rm ram}<10^{-8}$ inside the simulated ULX bubble, where $f_{\rm ram}$ is the ram pressure force. This result indicates that it is reasonable to neglect the radiative force compared to the ram pressure force. Although radiative force may become important in anisotropic systems or where the ram pressure force is weak, a detailed treatment is left for future work.

\subsection{Outflow injection} \label{sec:injection}

In this work, for simplicity, we only consider the disk outflows with a constant angular distribution of both the mass outflow rate $\dot{M}_{\rm w}$ and the momentum outflow rate $\dot{M}_{\rm w}v_0$, with a specific half opening angle. We put a stellar mass black hole at the center of the simulation box but its gravity is negligible at the distance of $\sim$pc. For simplicity, we also assume that the black hole is embedded in a stable and homogeneous ISM with a constant ISM number density and temperature.

We inject outflows with a specific opening angle by following the procedure below.  
\begin{enumerate}
\item We randomly select a target direction $\pm \boldsymbol{n}$ within the solid angle of the outflows as seen from the central black hole, using the Monte-Carlo method.  
\item For each cell, we calculate the angle $\theta_i$ between the selected target direction $\pm \boldsymbol{n}$ and the position vector $\boldsymbol{r}_i$ from the black hole, and identify all cells (called target cells) that satisfy $\theta_i < 3^\circ$ and lie inside the injection spheres.  
\item We then add mass and momentum to each target cell as
$\Delta M_i=\dot{M}_{\rm w}\Delta t_{\rm inj}/N$, $\Delta \boldsymbol{p}_i = \frac{\dot{M}_{\rm w}v_0\Delta t_{\rm inj}}{N}\frac{\boldsymbol{r}_i}{\left| \boldsymbol{r}_i \right|}$,
where $N$ is the number of the target cells. Here, $\Delta t_{\rm inj}$ is the injection timestep, which is independent of the simulation timestep, and we adopt $\Delta t_{\rm inj} = 50~{\rm yr}$ in this study. 
\end{enumerate}

\begin{figure}
\begin{center}
\includegraphics[width=1.0\columnwidth]{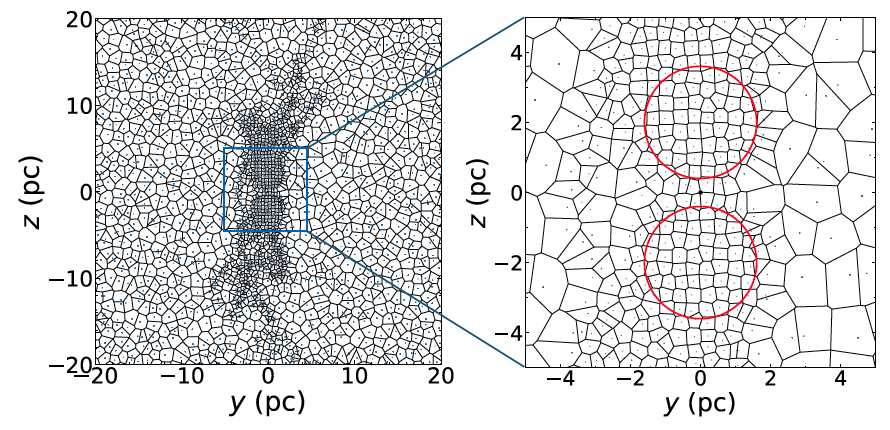}
\end{center}
\caption{Slice plots of the two hyper-refined injection regions on the plane that cuts through the north and south poles of the central black hole. The left panel shows the Voronoi mesh of the inner 20~pc of the simulation domain. The right panel shows the Voronoi mesh of the inner 5~pc. The red circles show the projections of the injection regions on the plane. \\
Alt text: two plots showing the Voronoi mesh of the simulation domain, and the zoom-in settings of the hyper-refined injection regions.
\label{fig:hyper_refine}}
\end{figure}

The \textit{injection spheres} are defined as two spheres centered at 2~pc away from the black hole along the north and south poles, respectively. Their radii are chosen so that the solid angle of each sphere as seen from the black hole is slightly larger than the solid angle of the disk outflows. For these two spheres, we set a target cell volume that is much smaller than that of the cells outside the injection regions. If a gas cell inside the injection regions has a volume larger (smaller) than twice (half) of the target volume, the cell is further refined (derefined). The limitation of the volume contrast between adjacent cells mentioned in Section~\ref{sec:setup} will prevent the cells inside the injection spheres from being immediately derefined after they leave the hyper-refined regions. Figure~\ref{fig:hyper_refine} shows the zoom-in view of the two hyper-refined injection regions \citep{Weinberger2017}, demonstrating that these regions can maintain the angular resolution to the central black hole during the simulation, and keep a smooth volume contrast to the large cells outside the regions. In addition, we inject outflows simultaneously in the target direction $\boldsymbol{n}$ in the northern hemisphere and in the opposite direction $-\boldsymbol{n}$ in the southern hemisphere, so that the momentum conservation of the injected outflows is guaranteed.  

For the fiducial run, we performed a test with a longer injection timestep of $\Delta t_{\rm inj} = 100~{\rm yr}$ and found no significant differences in the bubble structures. From this, we conclude that the simulation results have converged for $\Delta t_{\rm inj} = 50~{\rm yr}$. The frequency of the outflow injection will directly influence the computational cost of our simulations. This is because during the Monte-Carlo sampling process, we need to search for the target cells that satisfy the target direction. Such searching process costs about 58\% of the total CPU resource. As a result, we need to adopt a large $\Delta t_{\rm inj} \gg \Delta t_{\rm HD}$ to reduce the injection frequency and save computational resource, where $\Delta t_{\rm HD}$ is the hydrodynamical timestep decided by the Courant-Friedrichs-Lewy (CFL) condition. We find that the added mass per injection timestep in this condition will be larger than the pre-existing mass in one cell by at least one order of magnitude, which prevents the simulation runs from operating in an under-resolved regime. However, large $\Delta t_{\rm inj}$ will create some unphysical bow shock structures in the vicinity of the central black hole ($\sim$5~pc) because of the intermittent injection procedure, which is the main limitation of this Monte-Carlo method. Because we are focusing on the bubble structures at $\sim$100~pc scale, such small bow shocks will not influence our conclusions.

We note that although we adopt steady outflows in our simulations, the outflow rate and even the half opening angle can be variable in reality because of the change in mass accretion rate during ULX lifetime, e.g. the outflow rate will increase with a narrower opening angle when the accretion rate is higher. The influences of such variable wind properties to the bubble structures require further investigation, which we leave for a future work. However, our analysis of the wind inflated bubble nebulae can reveal the average properties of the ULXs during their $\sim$Myr lifetime.

\subsection{Parameter space} \label{sec:parameter}

In our simulation runs, we assume that the central black hole resides in a uniform ISM environment with a constant number density $n_{\rm ISM}$ and a constant temperature $T_{\rm ISM}=10^4$~K. We assume the solar abundance $Z_{\sun}$ for both the ISM and the injected outflows. We organize the simulation runs into three categories. The runs in the first category, called the "wind runs", adopt a half opening angle of $\alpha=45^{\circ}$, which correspond to the case of the wide-angle disk winds. The runs in the second category, called the "jet runs", adopt a half opening angle of $\alpha=5^{\circ}$, which correspond to the case of narrow funnel jets. For each of these categories, we vary the flow velocity $v_0$, the total mechanical power of the flow $L_{\rm mec}$, and the density of the ISM number $n_{\rm ISM}$, respectively, compared to the fiducial parameters. In the last category, we keep all the other parameters constant and only vary the half opening angle $\alpha$ of the outflows, to analyze the influence of the outflow angular distributions upon the bubble structures. We stop a simulation run once the long axis of the bubble has exceeded 90\% of the simulation box size, or the refinement process grows too computational expensive (e.g. the run BUB-25DEG) that it takes more than 10~hr CPU time to produce a new snapshot. The refinement process is related to the computational cost because of the additional computational cost of mesh refinement in the buffer zone between the hyper-refined spheres and the outer simulation domain. For large opening angle simulations, the volume contrast of hyper-refined regions and outer regions is relatively small; while for small opening angle simulations, the number of cells inside the buffer zone is small. Thus the run which costs most computational resource happens at a medium half opening angle about 20 deg.
The parameter settings of each simulation run are shown in Table~\ref{tab:initial}.

\begin{table*}
  \tbl{Initial simulation parameters.}
  {
  \begin{tabular}{cccccccc}
      \hline
      run name & $\alpha$ & $v_0$ & $L_{\rm mec}$ (\ergs) & $n_{\rm ISM}$ (cm$^{-3}$) & $T_{\rm ISM}$ (K) & $t_{\rm tot}$ (kyr) & $M_{\rm target}$ ($M_{\odot}$)\\
      (1) & (2) & (3) & (4) & (5) & (6) & (7) & (8)\\
      \hline
      \multicolumn{7}{c}{wind runs}\\
      \hline
      WIND-FID & $45^{\circ}$ & 0.1$c$ & $10^{39}$ & 0.1 & 10$^4$ & 304 & $5.87\times10^{-3}$\\
      WIND-LOWV & $45^{\circ}$ & 0.01$c$ & $10^{39}$ & 0.1 & 10$^4$ & 308 & $5.87\times10^{-3}$\\
      WIND-LOWL & $45^{\circ}$ & 0.1$c$ & $10^{37}$ & 0.1 & 10$^4$ & 387 & $5.87\times10^{-3}$\\
      WIND-LOWN & $45^{\circ}$ & 0.1$c$ & $10^{39}$ & 0.01 & 10$^4$ & 144 & $5.87\times10^{-4}$\\
      \hline
      \multicolumn{7}{c}{jet runs}\\
      \hline
      JET-FID & $5^{\circ}$ & 0.1$c$ & $10^{39}$ & 0.1 & 10$^4$ & 160 & $5.87\times10^{-3}$\\
      JET-LOWV & $5^{\circ}$ & 0.01$c$ & $10^{39}$ & 0.1 & 10$^4$ & 92 & $5.87\times10^{-3}$\\
      JET-LOWL & $5^{\circ}$ & 0.1$c$ & $10^{37}$ & 0.1 & 10$^4$ & 317 & $5.87\times10^{-3}$\\
      JET-LOWN & $5^{\circ}$ & 0.1$c$ & $10^{39}$ & 0.01 & 10$^4$ & 57 & $5.87\times10^{-4}$\\
      \hline
      \multicolumn{7}{c}{others}\\
      \hline
      BUB-25DEG & $25^{\circ}$ & 0.1$c$ & $10^{39}$ & 0.1 & 10$^4$ & 143 & $5.87\times10^{-3}$\\
      BUB-65DEG & $65^{\circ}$ & 0.1$c$ & $10^{39}$ & 0.1 & 10$^4$ & 321 & $5.87\times10^{-3}$\\ 
      BUB-SPH & $90^{\circ}$ & 0.1$c$ & $10^{39}$ & 0.1 & 10$^4$ & 302 & $5.87\times10^{-3}$\\
      \hline
   \end{tabular}}\label{tab:initial}
\begin{tabnote}
Column~1: Run name. \\
Column~2: Half opening angle of the outflow cone. \\
Column~3: Initial velocity of the outflows.  \\
Column~4: Total mechanical power of the outflows. \\
Column~5: Background ISM number density.\\
Column~6: Background ISM temperature. \\ 
Column~7: Total elapsed time of the run. \\
Column~8: Target cell mass of the run. \\
\end{tabnote}
\end{table*}

\section{Results} \label{sec:result}

\subsection{Wind bubble} \label{sec:wind_bubble}

\subsubsection{Bubble structures} \label{sec:wind_bubble_structure}

\begin{figure*}
\begin{center}
\includegraphics[width=0.9\textwidth]{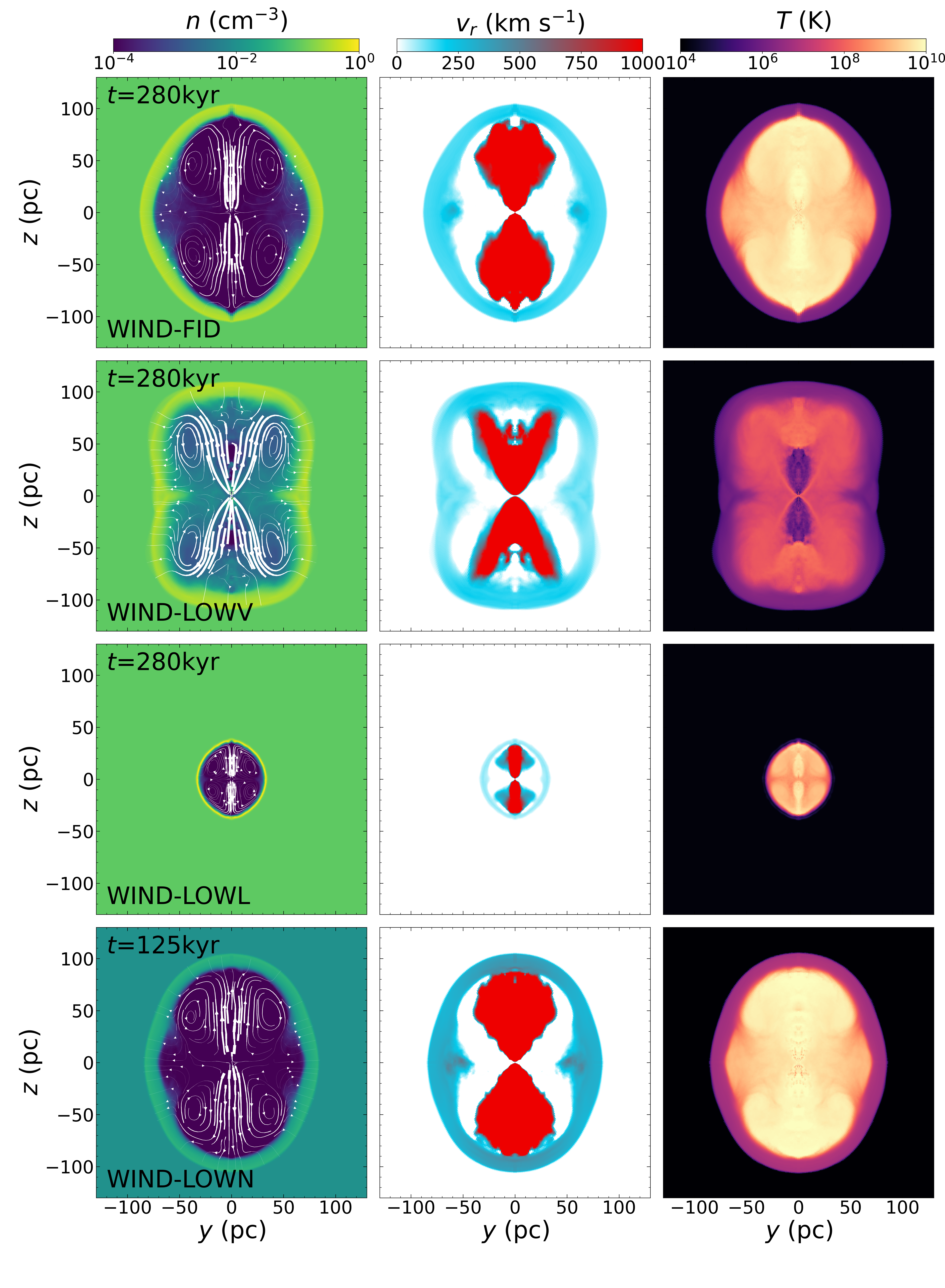}
\end{center}
\caption{Azimuthally averaged snapshots of the bubbles produced by the $45^{\circ}$ wide-angle disk winds. The left panels show the number density color maps overlaid with velocity streamlines, where the thickness of the streamlines represents the magnitude. The middle column panels and the right column panels show the color maps of the radial velocity and the gas temperature, respectively. From the top to the bottom row, the data come from the runs WIND-FID, WIND-LOWV, WIND-LOWL, and WIND-LOWN, respectively. All the snapshots are taken at $t=280$~kyr, except for the run WIND-LOWN, where the snapshot is taken at $t=125$~kyr.\\
Alt text: snapshots of the 4 simulation runs of wide opening angle disk winds, showing the number density, radial velocity, and gas temperature distributions.
\label{fig:wind_bubble}}
\end{figure*}

We show the wind bubble structures with the snapshots of the number density $n$ overlaid by velocity streamlines, radial velocity $v_r$, and the gas temperature $T$ in Figure~\ref{fig:wind_bubble}. We calculate the azimuthally averaged number density as $\left< n \right> = \frac{1}{2\pi} \int^{2\pi}_{0} n d\phi$. For the other properties, we calculate the number density weighted azimuthally averaged value as $\left< a \right>_{n} = \frac{\int^{2\pi}_{0} na d\phi}{\int^{2\pi}_{0} n d\phi}$, where $a$ represents radial velocity $v_r$ or gas temperature $T$. In order to show the radial velocity structures clearly, we set an upper limit cut 1000~km~s$^{-1}$ in Figure~\ref{fig:wind_bubble}. For the runs WIND-FID, WIND-LOWL, and WIND-LOWL, we select the same snapshots at the age $t=280$~kyr. For the run WIND-LOWN, we select the snapshot at the age $t=125$~kyr so that the position of the shock front at the $z$ axis is the same as that of the run WIND-FID.

Figure~\ref{fig:wind_bubble} shows that the basic structures of the wind bubbles are similar. A conical high velocity funnel emerges around the rotational axis with $v_r>$1000~km~s$^{-1}$, which consists of the gas outflows from the central accretion engine before shock. The outflows quickly decelerate at the termination shock. The shocked ISM forms a dense gas shell, which blocks the outflows and makes part of the outflows return along the inner surface of the gas shell to the mid-plane. Eventually, except for the run WIND-LOWV, the accretion disk winds drive an elliptical bubble interior with low number density of $\sim10^{-4}$~cm$^{-3}$ and high gas temperature of $\sim10^{10}$~K, covered by a dense gas shell with the number density of $\sim0.1$~cm$^{-3}$ and the moderate gas temperature of $10^6$~K. The gas shell propagates with a velocity of about $\sim200$~km~s$^{-1}$, which is roughly consistent with the shock velocity of NGC~55~ULX-1 measured by [O III] $\lambda$5007/H$\beta$ line ratio \citep{Zhou2023}.

Comparing the snapshots of WIND-FID and WIND-LOWV, we show that the initial outflow velocity greatly influences the morphology of the wind bubbles. With a lower initial velocity but the same mechanical power compared with the fiducial values, the outflows of the run WIND-LOWV have higher momentum, which makes the ISM difficult to deflect the original propagation of the outflows. As a result, the outflows with higher momentum tend to have more transverse propagation due to the transverse component of initial velocity, which makes the bubble of the run WIND-LOWV resemble a cylinder instead of an ellipsoid.

On the contrary, comparing the snapshots of the run WIND-FID and the run WIND-LOWL at the same age, it is found that the mechanical power of the outflows only influences the size of the bubble without changing its shape. We calculate the edge-on eccentricity $e_{90}$ of the bubble, where "edge-on" here means edge-on view of the central accretion disk, using the long axis $d_{\rm long}$ and the short axis $d_{\rm short}$ of the bubble as $e_{90}=\frac{\sqrt{d_{\rm long}^2-d_{\rm short}^2}}{d_{\rm long}}$, and summarize the results in Table~\ref{tab:morphology}. The bubble eccentricities of the runs WIND-FID and WIND-LOWL indeed only show a marginal variation from 0.53 to 0.43.

There is a degeneracy between the outflow mechanical power and the age of the system. If the age of the system is large enough, the outflows with low mechanical power may create a bubble that has the same size and shape as the bubble of the fiducial parameters. However, we argue that this degeneracy can be distinguished by the thickness of the gas shell. The propagation of the wind bubble will leave the adiabatic phase when the system age surpasses the cooling timescale. After the adiabatic propagation, the bubble shell will enter a snowplow phase and collapse because of radiative cooling \citep{Castor1975}. In the run WIND-FID, the cooling timescale of the gas shell is about 700~kyr, obtained from the simulation data by taking the mass-weighted average of $t_{\rm cool}=U_{\rm thermal}/Q^-_{\rm cool}$ over cells exceeding the background number density, where $U_{\rm thermal}$ is the cell’s total thermal energy and $Q^-_{\rm cool}$ is the cooling rate according to the cooling function. This value is consistent with the absence of shell collapse at $t=280$ kyr. For the run WIND-LOWL, the thin shell is not well resolved, making it difficult to derive the cooling timescale directly from the simulation data. Under the isotropic condition, however, the cooling timescale can be estimated as $t_{\rm cool}\geq1.7\times10^3\left( \dot{M}_6v_{2000}^2/n_{\rm ISM} \right)^{1/2}$~yr \citep{Castor1975}, where $\dot{M}_6 = \dot{M}_{\rm w}/(10^{-6}M_{\sun}~{\rm yr}^{-1})$, $v_{2000}=v_0/(2000~{\rm km}~{\rm s}^{-1})$. This yields a cooling timescale of $\sim$ 15 kyr, which is much shorter than  $t=280$~kyr shown in Figure 2. Therefore, the thinning of the shell in this run can be attributed to its collapse caused by radiative cooling. It is predictable that even when the bubble with low mechanical power grows to the same size as that of the fiducial bubble, the gas shell will be significantly thinner than that of the fiducial bubble, because of its much shorter cooling timescale.

The run WIND-LOWN shows that the background ISM density mainly influences the propagation speed of the gas shell. With the ISM density decreasing by one order of magnitude, the propagation speed of the gas shell is about twice the fiducial value. As a result, the bubble can grow to the same size as that of the fiducial run in a much shorter lifetime. The background ISM density will also influence the free expansion radius and the cooling timescale of the system. A lower background ISM density is expected to lead to a larger free expansion radius and a larger cooling timescale. We need another small scale simulation and another long elapsed time simulation to show the free expansion phase and the shell collapse at the cooling timescale, respectively, which is left for further works. However, our conclusions still hold for the adiabatic phase of the ULX bubble expansion.

The different initial ISM temperatures mainly lead to different outer thermal pressures against shock propagation. However, the sum of the inner thermal pressure and the ram pressure is more than 3 orders of magnitude larger than the ISM thermal pressure. The ISM thermal pressure will only be important when the total inner pressure reduces to the same magnitude. From the WIND-FID run, we calculate the timescale for the total pressure at the gas shell to reduce to the magnitude of the ISM thermal pressure. The timescale is about 2~Myr, which is much longer than the elapsed time of our simulation. Thus, we conclude here that the background ISM temperature only has a negligible influence on our simulation results. 

\begin{table*}
  \tbl{Morphology of the bubbles in all the simulation runs.}
  {
  \begin{tabular}{cccccc}
      \hline
      run name & shape & $d_{\rm long}$ (pc) & $d_{\rm short}$ (pc) & $e_{90}$ & $t$ (kyr)\footnotemark[$*$]\\
      (1) & (2) & (3) & (4) & (5) & (6)\\
      \hline
      \multicolumn{6}{c}{wind runs}\\
      \hline
      WIND-FID & ellipsoid & 214 & 181 & 0.53 & 280 \\
      WIND-LOWV & cylinder & 222 & 154 &  & 280 \\
      WIND-LOWL & ellipsoid & 82 & 74 & 0.43 & 280 \\
      WIND-LOWN & ellipsoid & 213 & 171 & 0.60 & 125 \\
      \hline
      \multicolumn{6}{c}{jet runs}\\
      \hline
      JET-FID & ellipsoid & 293 & 110 & 0.93 & 150\\
      JET-LOWV & ellipsoid & 290 & 33 & 0.99 & 73\\
      JET-LOWL & ellipsoid & 67 & 50 & 0.67 & 150\\
      JET-LOWN & ellipsoid & 289 & 77 & 0.96 & 45\\
      \hline
      \multicolumn{6}{c}{others}\\
      \hline
      BUB-25DEG & ellipsoid & 182 & 119 & 0.76 & 280\\
      BUB-65DEG & ellipsoid & 198 & 183 & 0.38 & 280\\ 
      BUB-SPH & sphere & 186 & 186 & 0.0 & 143\\
      \hline
   \end{tabular}}\label{tab:morphology}
\begin{tabnote}
Column~1: Run name. \\
Column~2: Shape of the bubble. \\
Column~3: Long axis of the bubble. \\
Column~4: Short axis of the bubble. \\ 
Column~5: Edge-on eccentricity of the bubble if its shape resembles an ellipsoid. \\
Column~6: Snapshot time. \\
\footnotemark[$*$] The snapshots are taken at the same time with those in Figure~{\ref{fig:wind_bubble}} and Figure~{\ref{fig:jet_bubble}}. The snapshots of the runs BUB-65DEG and BUB-SPH are taken at $t=280$~kyr in order to compare with the run WIND-FID. The snapshot of the run BUB-25DEG is taken at the end of the simulation due to the limitation of computational resource.
\end{tabnote}
\end{table*}

\subsubsection{Shock structures} \label{sec:wind_shock_structure}

\begin{figure}
\begin{center}
\includegraphics[width=\columnwidth]{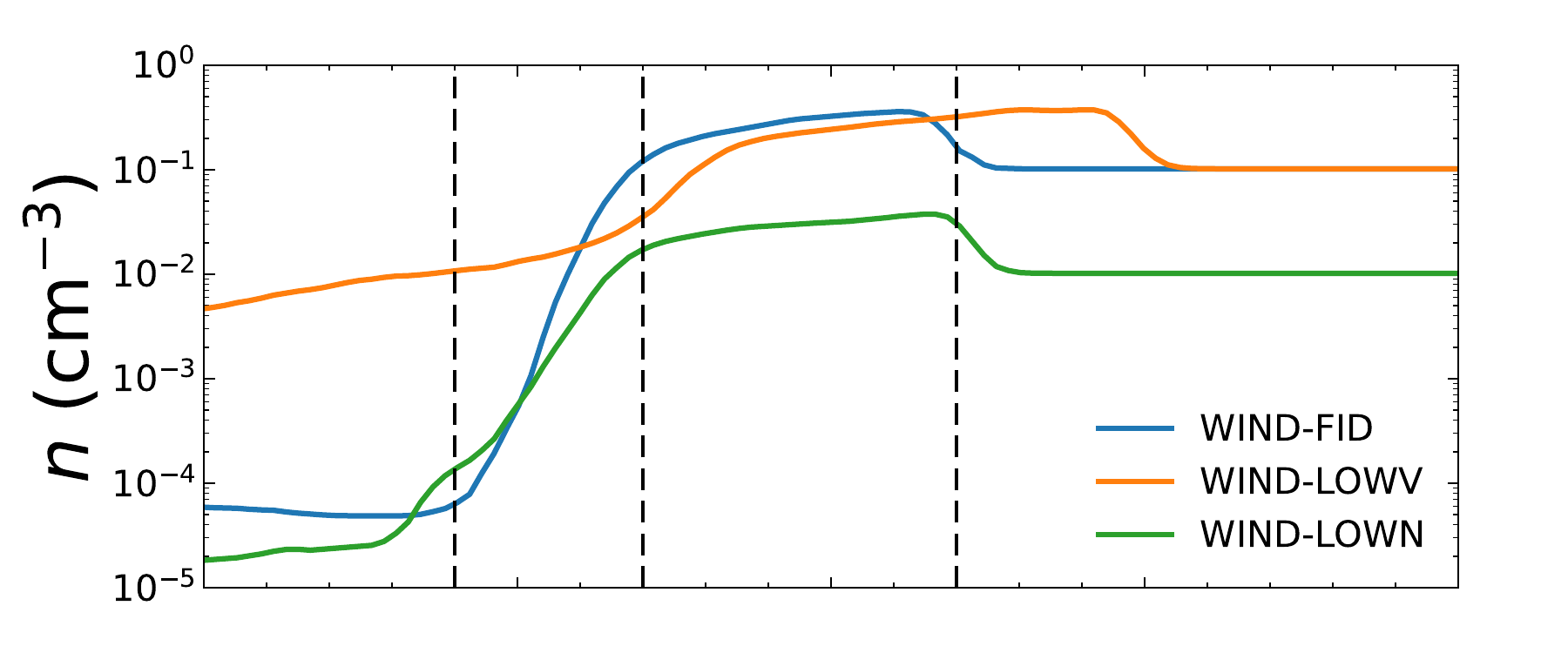}
\includegraphics[width=\columnwidth]{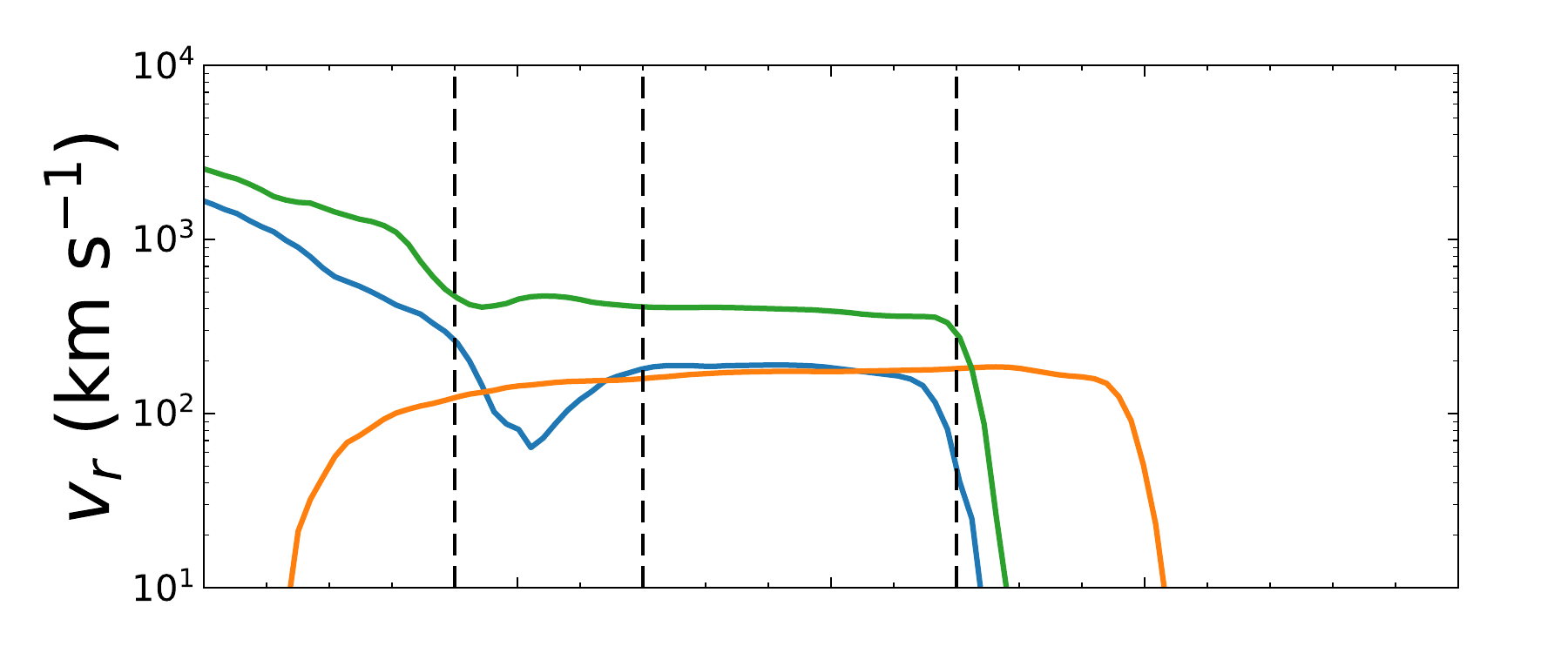}
\includegraphics[width=\columnwidth]{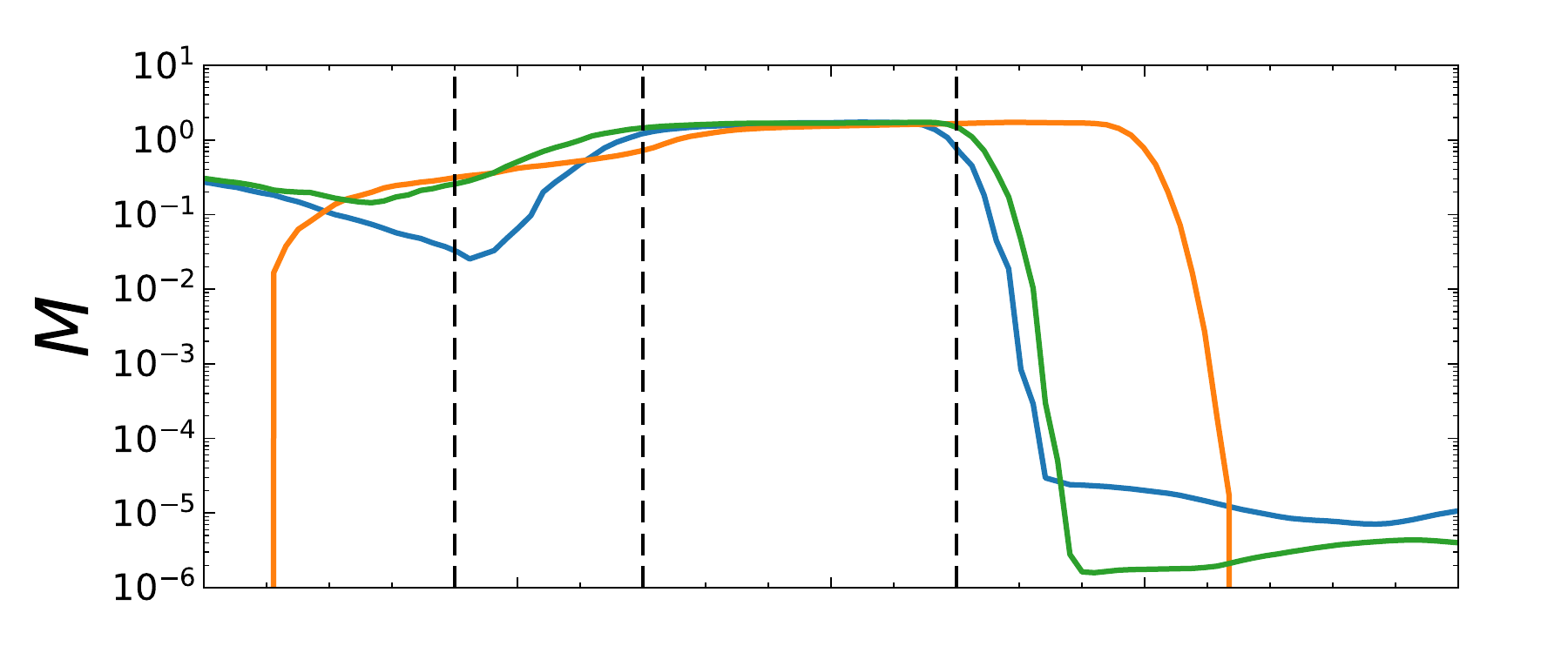}
\includegraphics[width=\columnwidth]{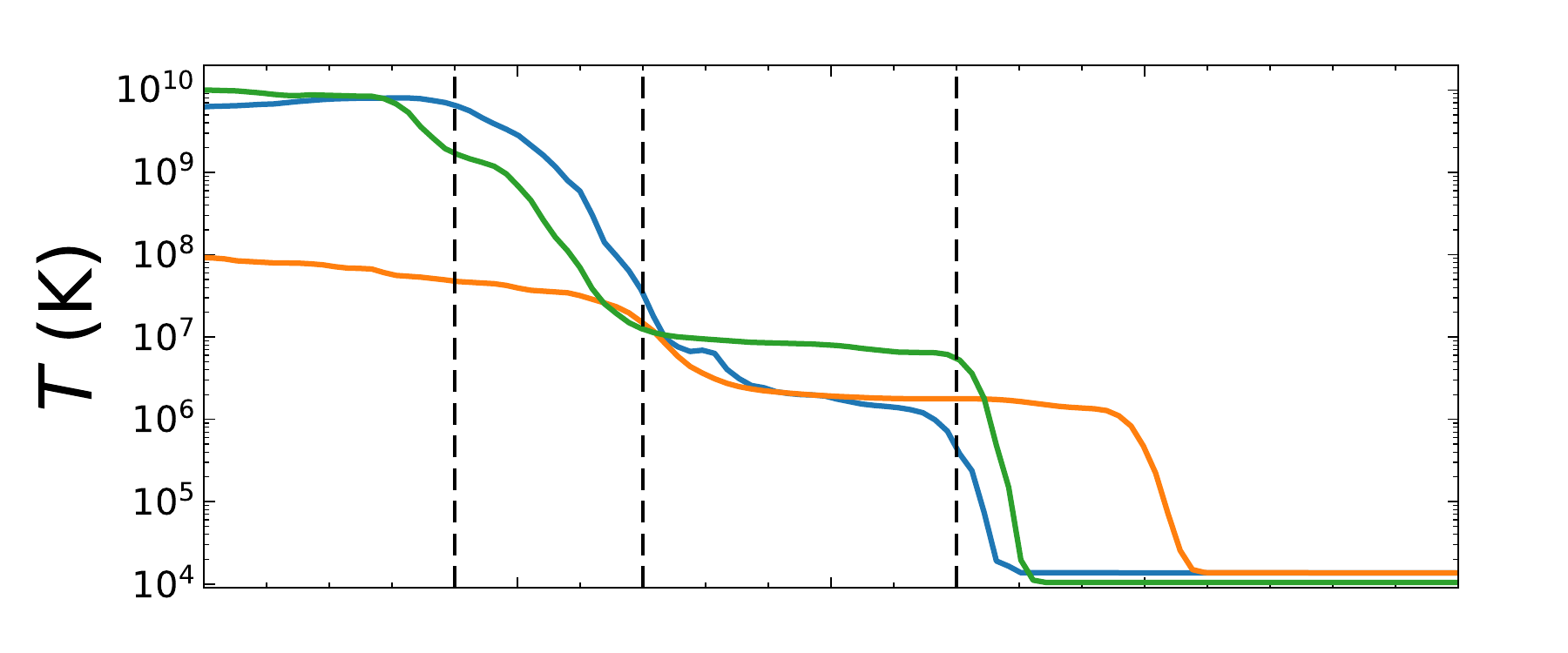}
\includegraphics[width=\columnwidth]{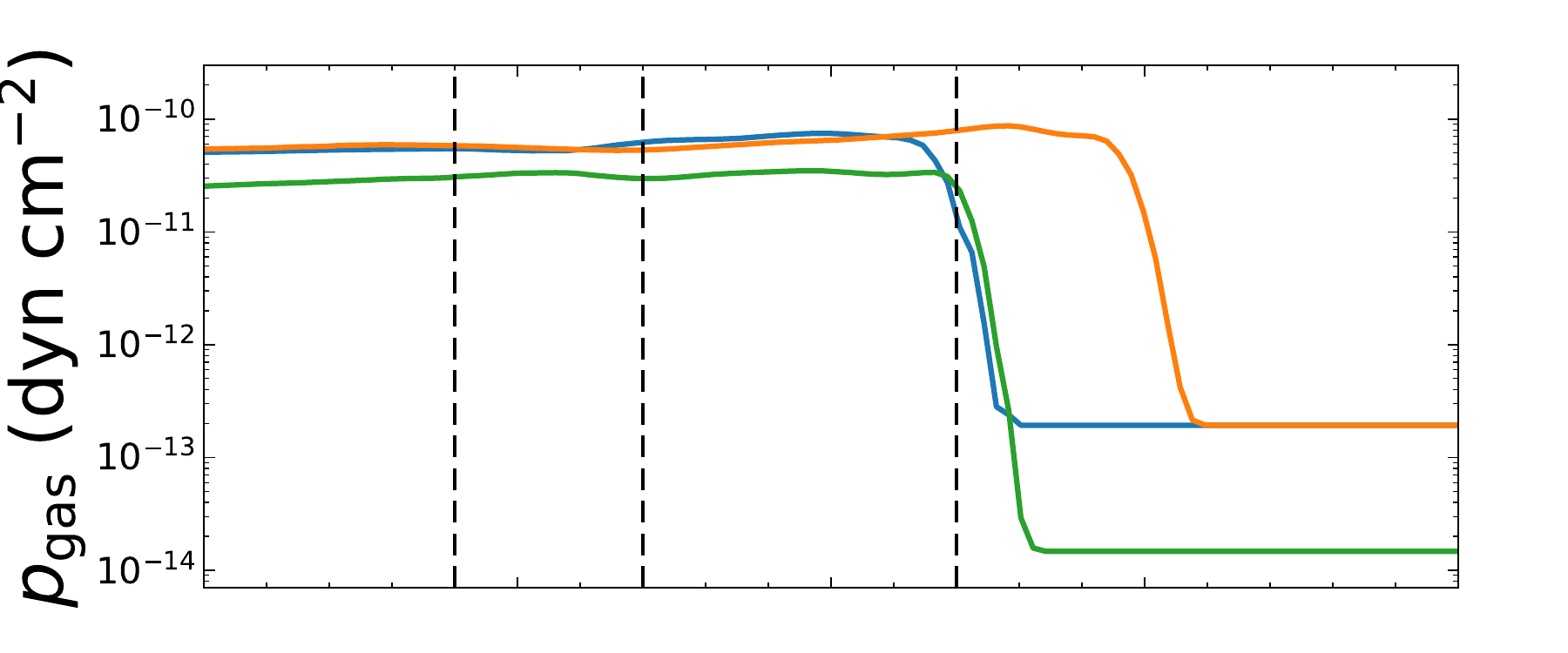}
\includegraphics[width=\columnwidth]{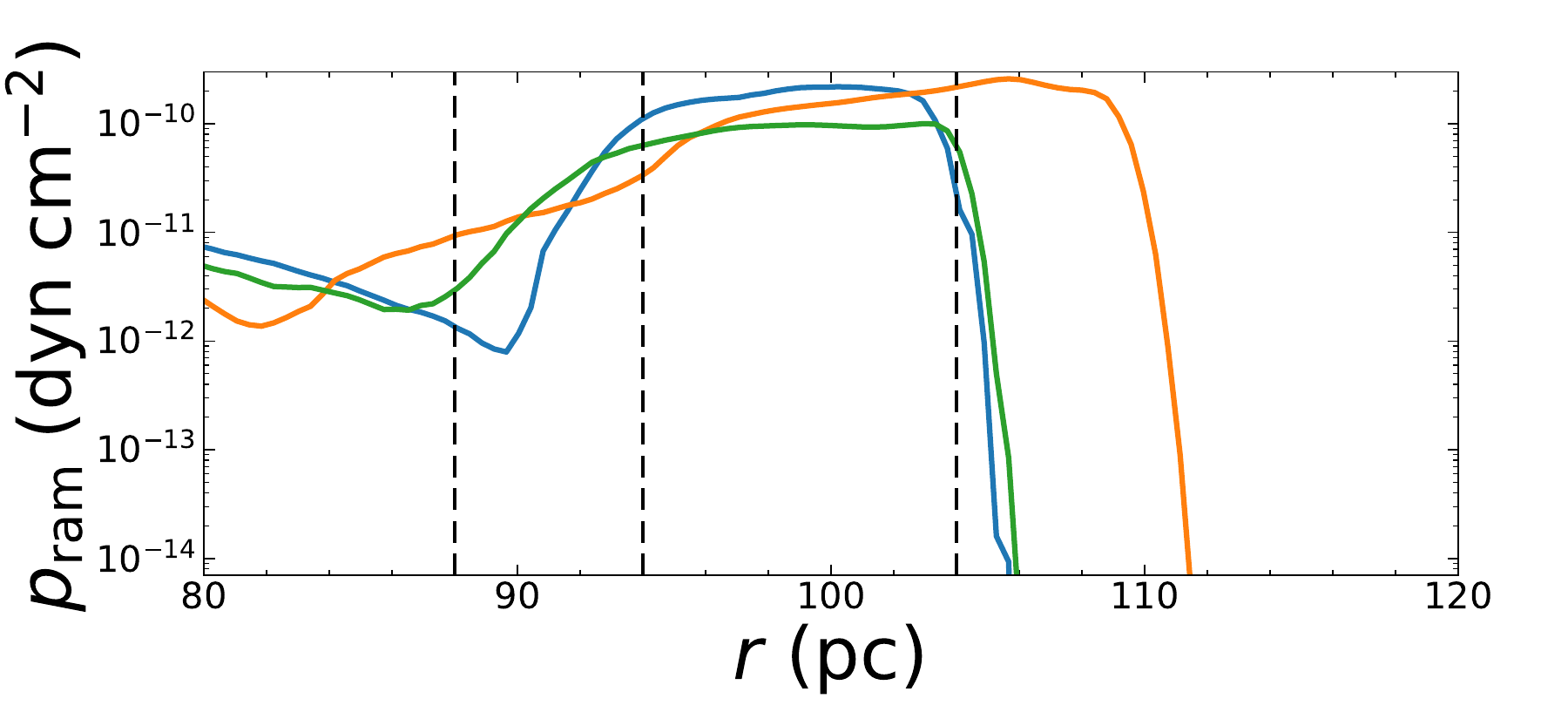}
\end{center}
\caption{Radial profiles of number density $n$, radial velocity $v_r$, local Mach number $\mathcal{M}$, gas temperature $T$, gas pressure $p_{\rm gas}$, and ram pressure $p_{\rm ram}$ of the runs WIND-FID, WIND-LOWV, and WIND-LOWN, respectively. All the radial profiles are taken from the same snapshots of Figure~\ref{fig:wind_bubble}. The data is azimuthally and pitch angle averaged between $\theta=0^{\circ}$ and $\theta=10^{\circ}$.\\
Alt text: radial profiles of shock structures of the 4 wind runs, showing the 1D structures of number density, radial velocity, local Mach number, temperature, gas pressure, and ram pressure, respectively.
\label{fig:wind_radial}}
\end{figure}

The shock produced by the disk wind will only dissipate into the background ISM when the inner pressure and the outer pressure balance. We have demonstrated that the elapsed time of our simulation is much shorter than this timescale in Section~\ref{sec:wind_bubble_structure}. As a result, we can analyze the unperturbed shock structures of the bubble shell. Figure~\ref{fig:wind_radial} shows the pitch angle averaged radial profiles of number density $n$, radial velocity $v_r$, gas temperature $T$, gas pressure $p_{\rm gas}$, and ram pressure $p_{\rm ram}$ for 3 of our wind simulation runs. We do not use the data of the run WIND-LOWL because the bubble size is too small to compare with the other 3 runs. 

In all models, the shock can be divided into 4 parts: the post shock disk wind (<88~pc), the contact discontinuity ($\sim$88--94~pc), the shocked ISM ($\sim$94--104~pc), and the unperturbed ISM (>104~pc), except for the run WIND-LOWV, where the forward shock moves a little outward. The post-shock disk wind area has a low number density and a high gas temperature. The number density and temperature of the post-shock wind strongly depend on the initial velocity but are not greatly influenced by the background ISM density. For the run WIND-LOWV, a lower initial disk wind velocity leads to higher inner number density and lower gas temperature of the wind bubble. This is because disk winds with a lower initial velocity have a higher mass load, and the density of the post shock wind is directly related to the density of outflows itself. In all models, the disk wind gradually decelerates before $\sim$90~pc. Between about 90--95~pc, the number density gradually increases, but the radial velocity and the gas pressure in this area remain constant. This is a typical characteristic for contact discontinuity. In the shocked ISM area, the properties are determined mainly by the initial ISM parameters, regardless of the wind velocity. The number density of shocked ISM is about 4 times the background ISM number density, so a lower initial ISM density in the run WIND-LOWN leads to a less dense bubble shell. The local Mach number $\mathcal{M}=v_r/c_{\rm s}$, where $c_{\rm s}$ is the local sound speed, reaches maximum in the shocked ISM region. The shocks are propagating at a similar local Mach number $\mathcal{M}\sim2$ in all these 3 runs. In addition, in the shocked ISM area, the gas thermal pressure and the ram pressure are comparable within an order of magnitude in all models, which indicates that the disk wind is already sufficiently thermalized.

The shock structures of our run WIND-LOWV are well consistent with those of the run L40-v2 in \citet{Siwek2017} before their radiative stage (see the red line of their Figure~2). We show a similar extended shocked ISM shell with 4 times the density of the background ISM and a temperature around $10^6$~K, which is also present in \citet{Siwek2017}. Meanwhile, they show a reverse shock at about 20~pc separating low-temperature free expansion winds and high-temperature shocked wind, which is also present at about 50~pc along the axis of our run WIND-LOWV in Figure~\ref{fig:wind_bubble} (see the free expansion winds with dark purple color and the shocked winds with dark red color in the temperature snapshot). However, the reverse shock is not present in the other runs of our simulation. This is because the thickness of the shocked wind layer will be larger if the wind velocity is larger (see Figure~6 in \citet{Siwek2017}). In the other runs with a higher wind velocity of 0.1~$c$, the shocked wind layer has already filled the entire bubble cavity, making the reverse shock disappear. One of the differences between our run WIND-LOWV and their run L40-v2 is that we show a much extended contact discontinuity in Figure~\ref{fig:wind_radial}, which instead is nearly a step function in the density and temperature curves of their work. This can be explained by the anisotropy of our wind injection. The distance of the contact discontinuity from the central source should depend on the pitch angle $\theta$. Figure~\ref{fig:wind_radial} is averaged between $\theta=0^{\circ}$ and $\theta=10^{\circ}$, which smooths out the different contact discontinuities between 88~pc and 94~pc. In addition, their work shows the radiative stage after the adiabatic expansion, when the bubble shell collapses. In contrast, we do not see such a stage in WIND-LOWV because we adopt a lower background ISM density thus a larger radiative cooling timescale. We left the simulation containing the radiative stage to a future work.

\begin{figure}
\begin{center}
\includegraphics[width=\columnwidth]{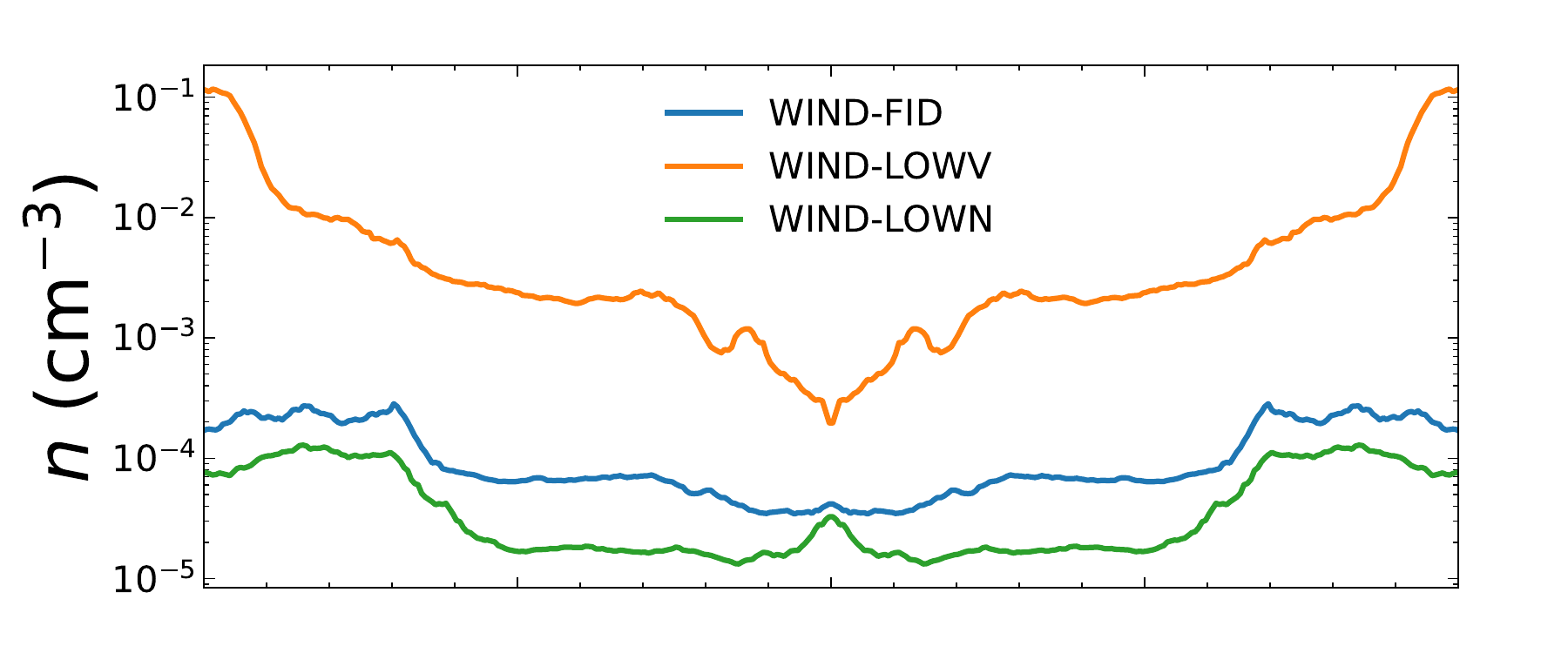}
\includegraphics[width=\columnwidth]{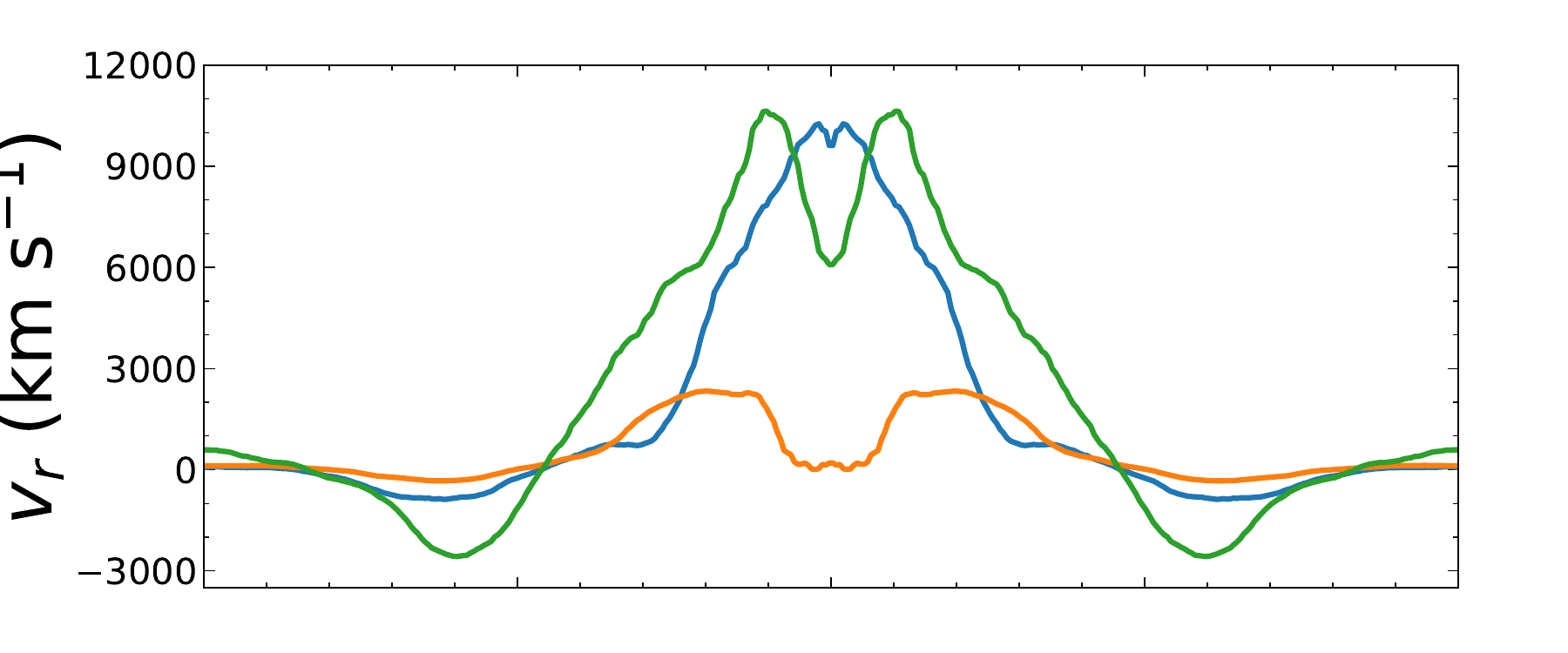}
\includegraphics[width=\columnwidth]{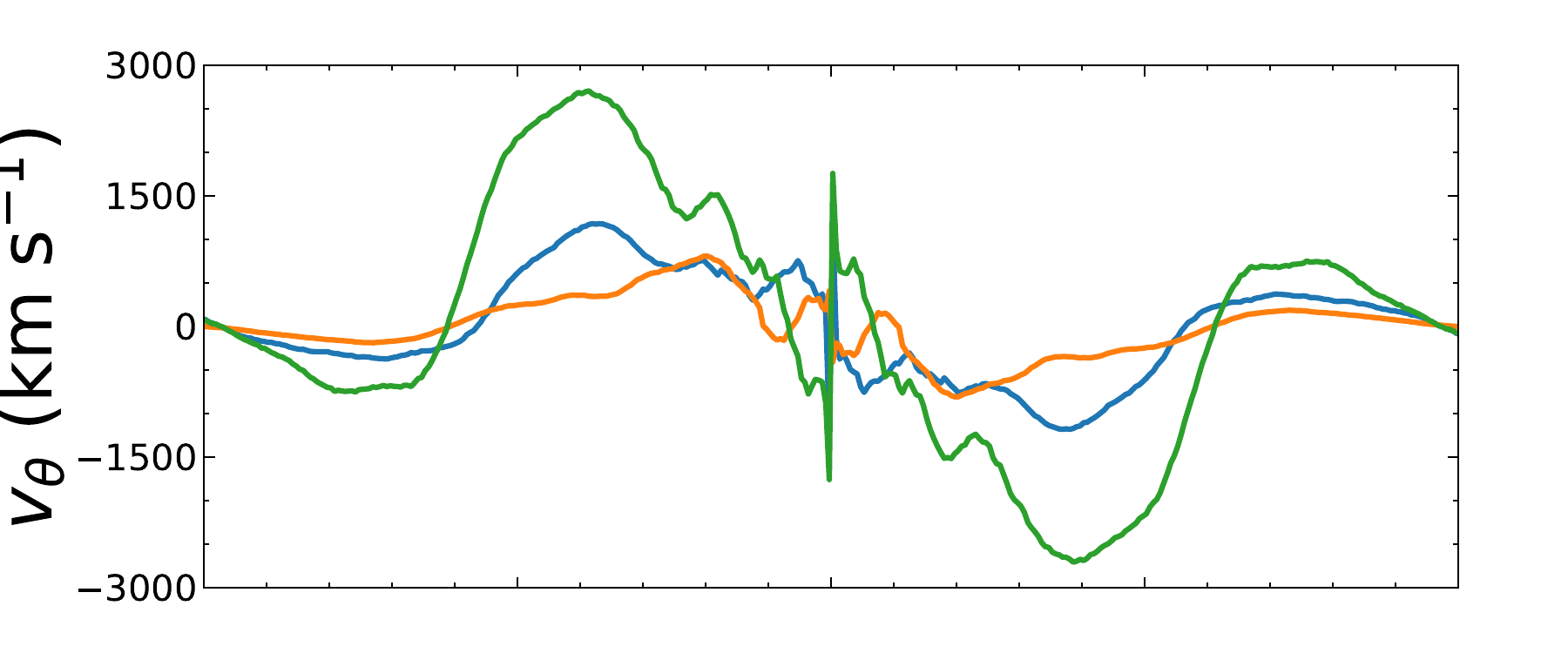}
\includegraphics[width=\columnwidth]{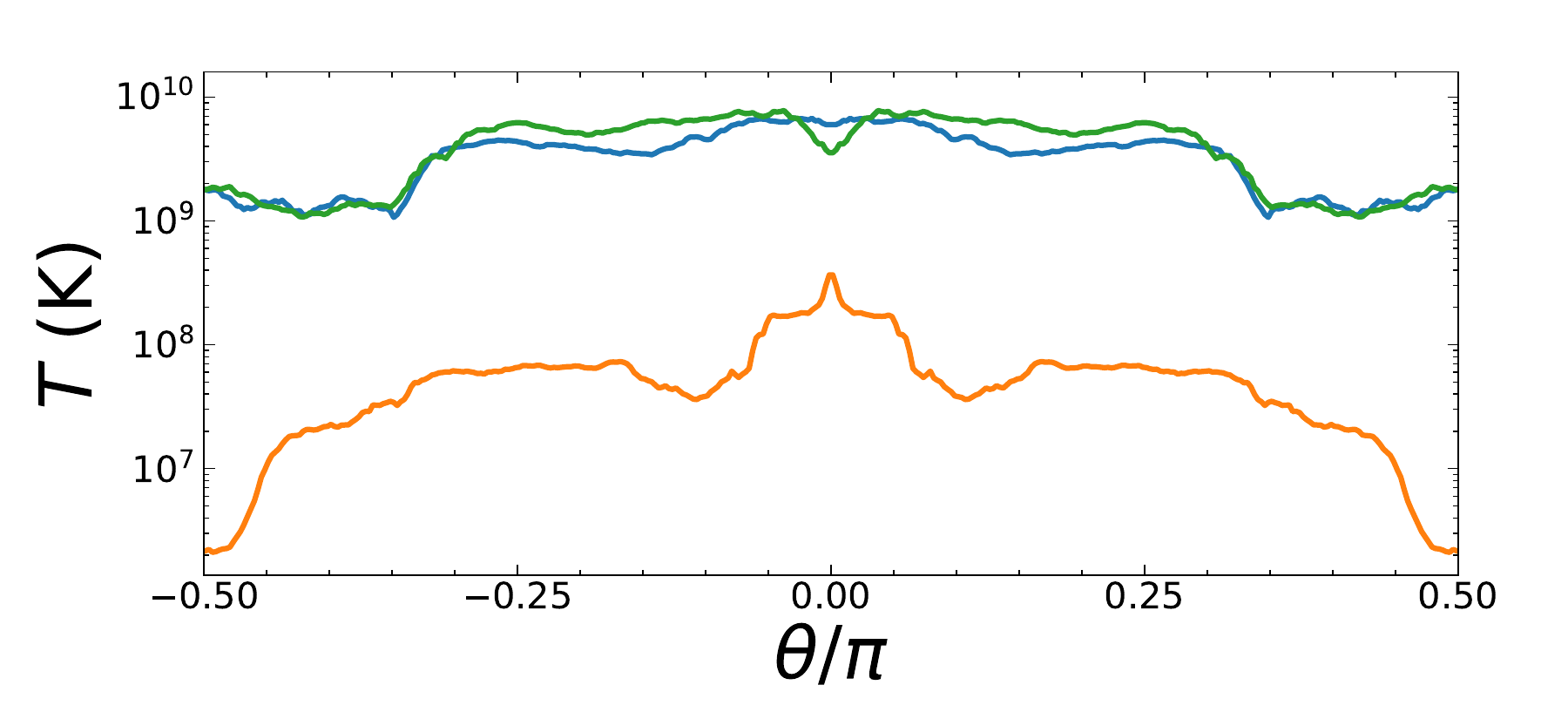}
\end{center}
\caption{Azimuthally averaged $\theta$ direction profiles of number density $n$, radial velocity $v_r$, transverse velocity $v_{\theta}$, and gas temperature $T$ of the runs WIND-FID, WIND-LOWV, and WIND-LOWN, respectively. All the transverse profiles are taken from the same snapshots of Figure~\ref{fig:wind_bubble} and at $r=50$~pc from the central black hole.\\
Alt text: transverse profiles of bubble structures of the 4 wind runs, showing the structures of number density, radial velocity, transverse velocity, and temperature, respectively.
\label{fig:wind_transverse}}
\end{figure}

In order to show the delicate structures of the bubble interior, we extract the $\theta$ direction profiles at 50~pc, of number density $n$, radial velocity $v_r$, transverse velocity $v_{\theta}$, and gas temperature $T$, and show in Figure~\ref{fig:wind_transverse}. For the two elliptical bubbles (WIND-FID and WIND-LOWN), the number density and temperature of the bubble interior show a clear jump at about $45^{\circ}$, which is exactly the half opening angle of the original disk wind. Inside the wind cone, the disk wind can directly push away the ISM and be more efficiently thermalized, so the number density is lower while the gas temperature is higher inside the cone. The radial velocity shows a sharp drop from the funnel to the edge of the wind cone. Meanwhile, the transverse velocity pointing to the axis shows a clear increase, which indicates the recollimation effect inside the wind cone. Outside the cone, the radial velocity is below 0, which shows the returning gas that we discussed in Section~\ref{sec:wind_bubble_structure}. For the cylindrical bubble of WIND-LOWV, a low density and nearly static structure appears at $\theta=0^{\circ}$. This can be seen in Figure~\ref{fig:wind_bubble} as the ``X''-shaped structure in the radial velocity distribution, which is a small part of the ISM that is trapped by the surrounding high velocity outflows. This structure can be potentially caused by the Monte-Carlo method we described in Section~\ref{sec:methods}. If this is the case, further improvement of the outflow injection method is needed in our future works.

\subsection{Jet bubble} \label{sec:jet_bubble}

\subsubsection{Bubble structures} \label{sec:jet_bubble_structure}

\begin{figure*}
\begin{center}
\includegraphics[width=0.9\textwidth]{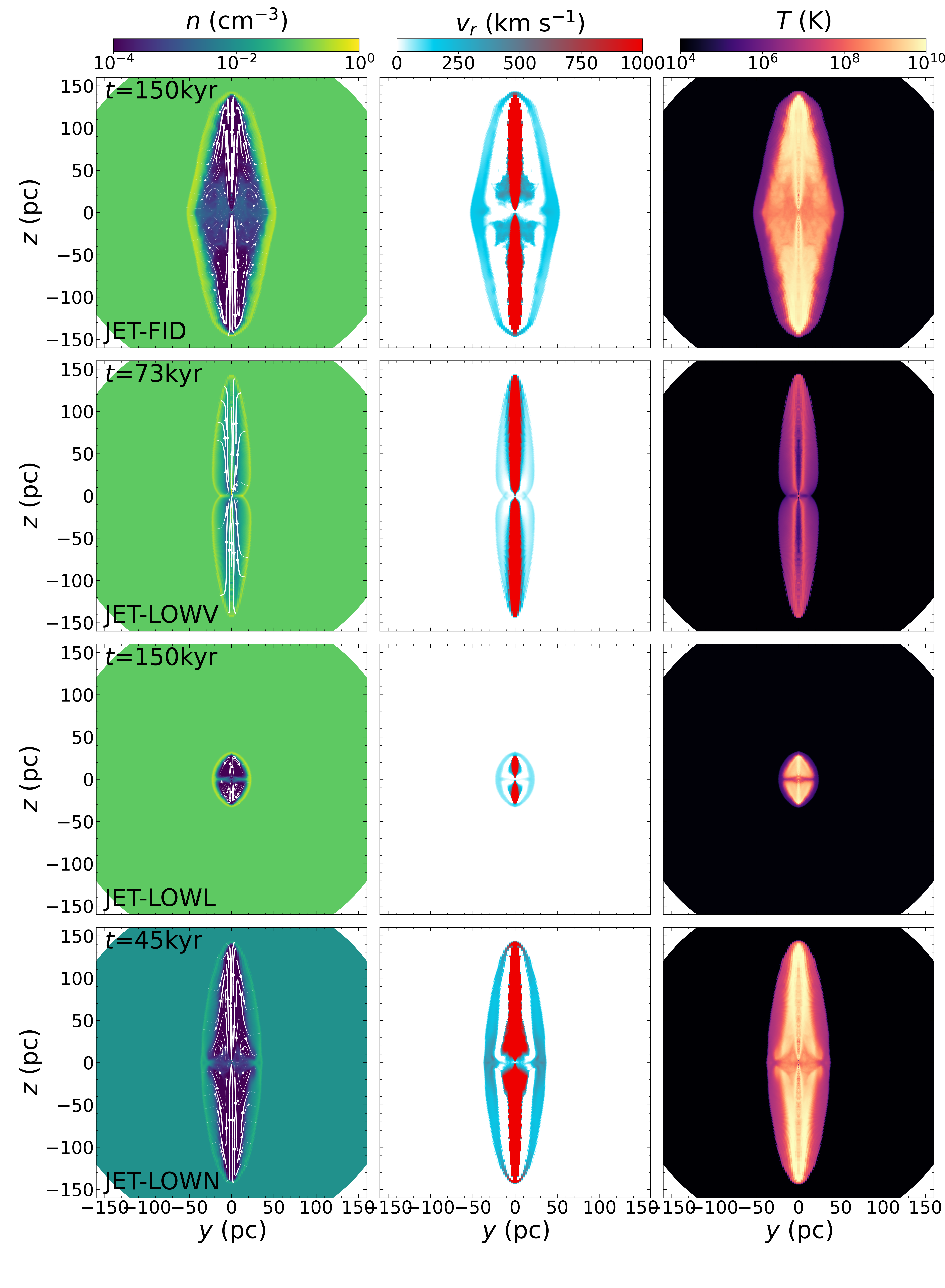}
\end{center}
\caption{Azimuthally averaged snapshots of the bubbles produced by the $5^{\circ}$ narrow-angle jets. The left panels show the number density color maps overlaid with velocity streamlines, where the thickness of the streamlines represents the magnitude. The middle column panels and the right column panels show the color maps of the radial velocity and the gas temperature, respectively. From the top to the bottom row, the data come from the runs JET-FID, JET-LOWV, JET-LOWL, and JET-LOWN, respectively. The snapshots of JET-FID and JET-LOWL are taken at $t=150$~kyr. The snapshots of the runs JET-LOWV and JET-LOWN are taken at $t=73$~kyr and $t=45$~kyr, respectively, when the shock front is at the same distance as that of JET-FID.\\
Alt text: snapshots of the 4 simulation runs of narrow opening angle jets, showing the number density, radial velocity, and gas temperature distributions.
\label{fig:jet_bubble}}
\end{figure*}

We show the jet bubble structures with the snapshots of the number density $n$ overlaid by velocity streamlines, radial velocity $v_r$, and the gas temperature $T$ in Figure~\ref{fig:jet_bubble}. The data is azimuthally averaged with the same method as described in Section~\ref{sec:wind_bubble_structure}. For the runs JET-FID and JET-LOWL, we select the same snapshots at the age $t=150$~kyr. For the runs WIND-LOWL and WIND-LOWN, we select the snapshots at the age $t=73$~kyr and $t=45$~kyr, so that the positions of the shock fronts at the $z$ axis are the same as that of the run WIND-FID. The eccentricities of the jet bubbles are also summarized in Table~\ref{tab:morphology}.

The overall structures of the jet driven bubbles are mostly similar with those of the wind driven bubbles. Taking JET-FID as an example, we can still see a high velocity outflow funnel ($>1000$~km~s$^{-1}$) near the system axis, which is shown as the yellow part in the $v_r$ snapshot. In addition, the entire bubble can be divided into a cavity with low number density (dark blue part in the $n$ snapshot) and high gas temperature (yellow part in the $T$ snapshot) inside the bubble, and a high number density (light green part in the $n$ snapshot), warm gas shell (purple part in the $T$ snapshot) surrounding the cavity. However, in contrast to the wind bubbles, the high velocity funnel of the jet bubbles is more collimated. According to the velocity streamlines, the recollimation is due to the returning gas, which pushes the outflowing gas from all around the funnel. A new feature of the jet bubbles is a low gas temperature jet core at the vicinity of the jet axis, shown as the orange part in the $T$ snapshot of JET-FID, which extends to about 100~pc before it is disrupted by turbulence. The jet core consists of the gas directly coming from the central accretion engine, which has a lower temperature than that of the bubble cavity because it is not sufficiently thermalized. We note that these trends hold for the other 3 runs JET-LOWV, JET-LOWL, and JET-LOWN, although they are not as obvious as the run JET-FID.

\begin{figure}
\begin{center}
\includegraphics[width=0.9\columnwidth]{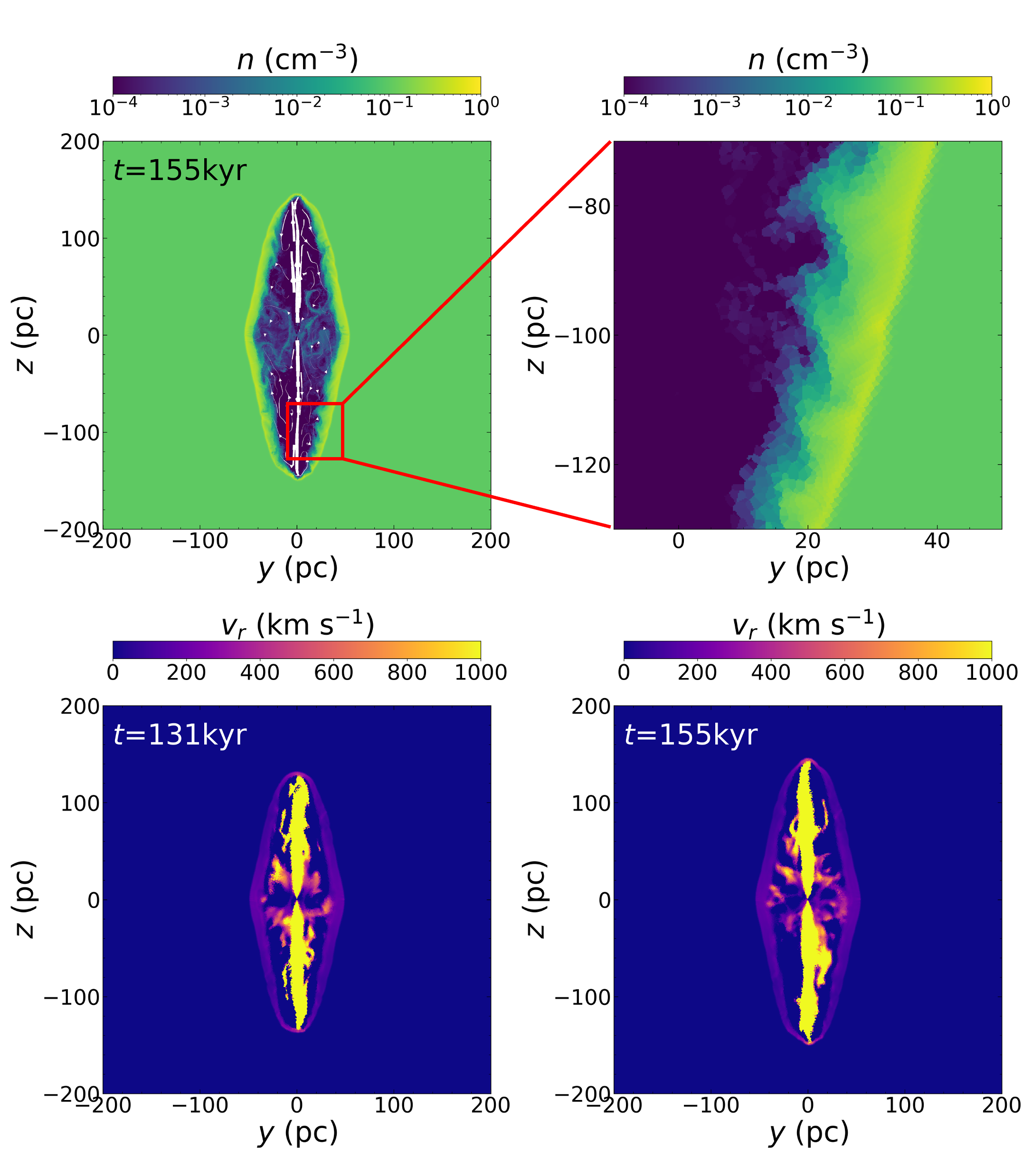}
\end{center}
\caption{{\bf Top row:} Slice plots of the number density $n$ of the run JET-FID at the snapshot $t=$155~kyr. The top right panel is a zoom-in view of the top left panel. The zoom-in view shows clear Kelvin-Helmholtz instability at the contact discontinuity. {\bf Bottom row:} Slice plots of the radial velocity $v_r$ of the run JET-FID at the snapshots $t=$131~kyr and $t=$155~kyr, respectively. The high velocity jet above the central black hole is wobbling after $\sim$80~pc, which resembles the Kink instability.\\
Alt text: snapshots of the fiducial jet run, showing the Kelvin-Helmholtz instability (top) and the wobbling instability (bottom).
\label{fig:instability}}
\end{figure}

We see obvious instabilities in the run JET-FID and show them in Figure~\ref{fig:instability}. In the top row, a clear Kelvin-Helmholtz (KH) instability pattern with a wavelength of $\sim$10~pc is present at the contact discontinuity. In addition, the growth timescale for the KH instability is $\tau_{\rm KH}=\frac{\rho_1 + \rho_2}{\left( \rho_1 \rho_2 \right)^{0.5}} \frac{\lambda}{2 \pi \Delta v}$ \citep{Chandrasekhar1961}, where $\rho_1$ and $\rho_2$ are the densities of both sides across the contact discontinuity, $\lambda$ is the wavelength of the instability, $\Delta v$ is the relative velocity of the shear flow. The KH instability with a wavelength of 10~pc can grow with a timescale of $\sim$100~kyr, which is comparable to the simulation elapsed time of JET-FID. The corresponding growth timescale for the run JET-LOWV is $\sim$1~Myr because of lower shear velocity, which is much longer than the elapsed time of the run. Thus, we cannot see the KH instability pattern but a smooth contact discontinuity in the run JET-LOWV. In the bottom row, we see the jet propagation direction is wobbling beyond $\sim$80~pc, which resembles the electromagnetic Kink instability. However, we do not include magnetic fields in this work. This instability is a purely hydrodynamic effect due to the asymmetry of the returning gas. These instabilities disrupt the jet and thermalize it, which terminates the low temperature jet core at about 100~pc.

Comparing different jet runs in Figure~\ref{fig:jet_bubble} and Table~\ref{tab:morphology}, we can still conclude that the initial momentum of the jet alters the eccentricity of the bubble, and the mechanical power of the jet alters the bubble size. However, unlike the conclusions of the wind runs, the initial momentum of the jet also influences the propagation speed of the termination shock front. This is because the jet with high initial momentum is not completely thermalized at the termination shock, as seen in the temperature snapshot of the run JET-LOWV. Thus, the propagation speed of the shock front still greatly depends on the initial velocity of the jet. In addition, reducing the mechanical power by 2 orders of magnitude from JET-FID to JET-LOWL also results in a significant decrease of eccentricity from 0.93 to 0.67. From the temperature snapshot of JET-LOWL, the jet is quickly thermalized because of its low mechanical power, as a result the propagation of the bubble in JET-LOWL is more isotropic than that of the fiducial jet bubble JET-FID. The influence of background ISM is still similar to the conclusions of the wind runs.

\subsubsection{Shock structures} \label{sec:jet_shock_structure}

\begin{figure}
\begin{center}
\includegraphics[width=\columnwidth]{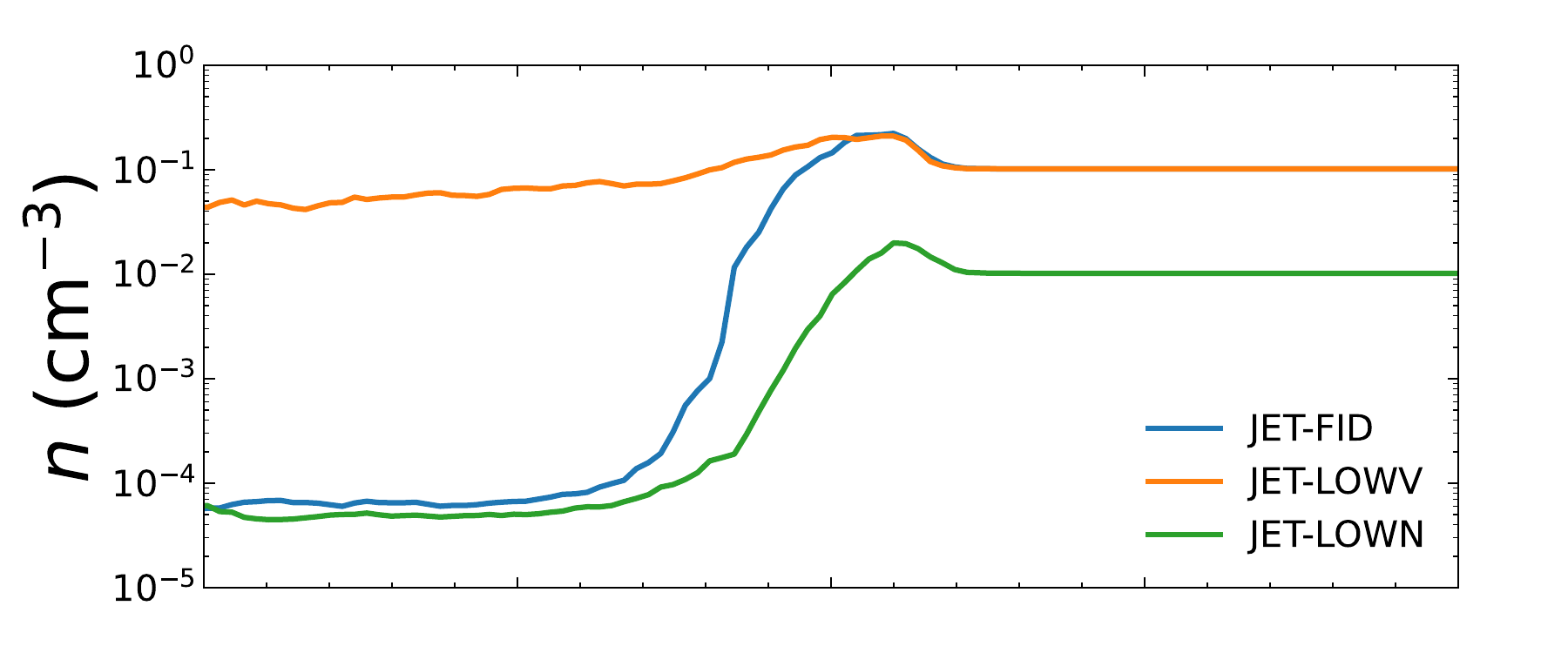}
\includegraphics[width=\columnwidth]{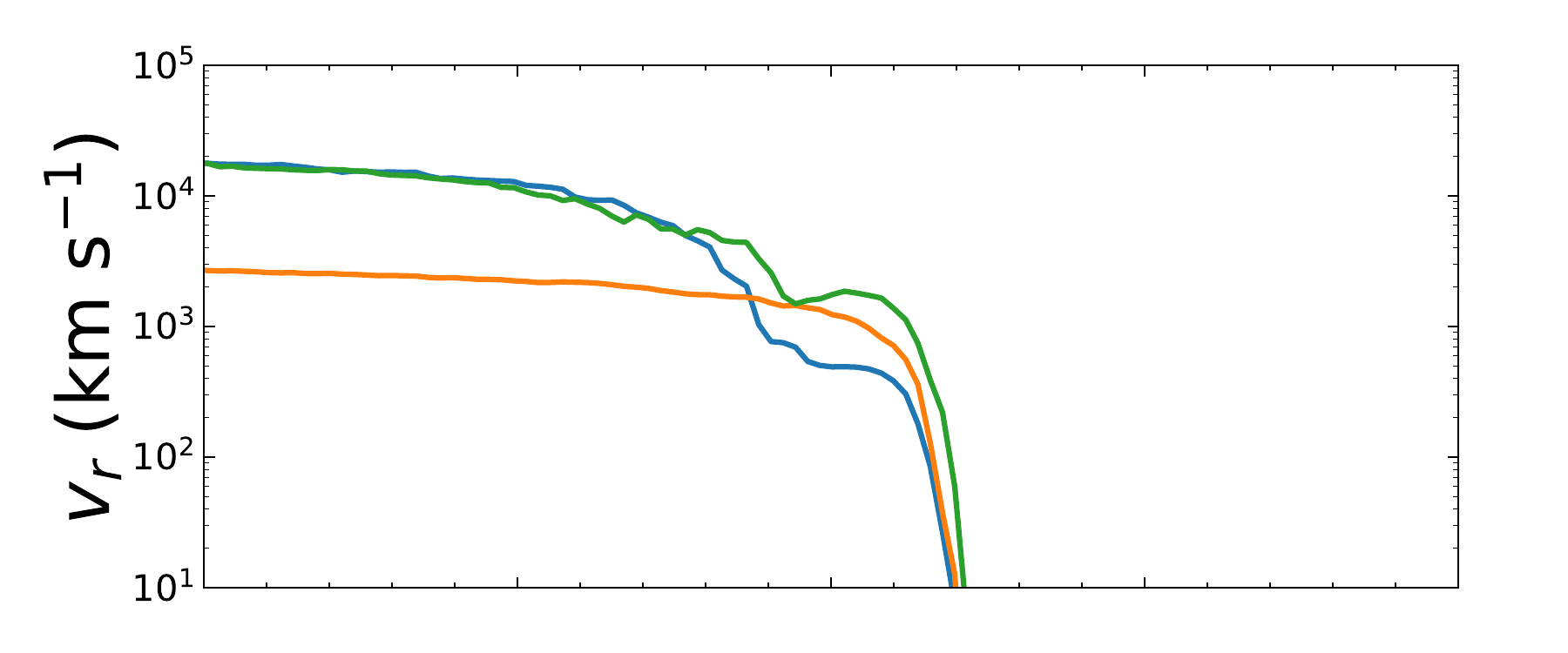}
\includegraphics[width=\columnwidth]{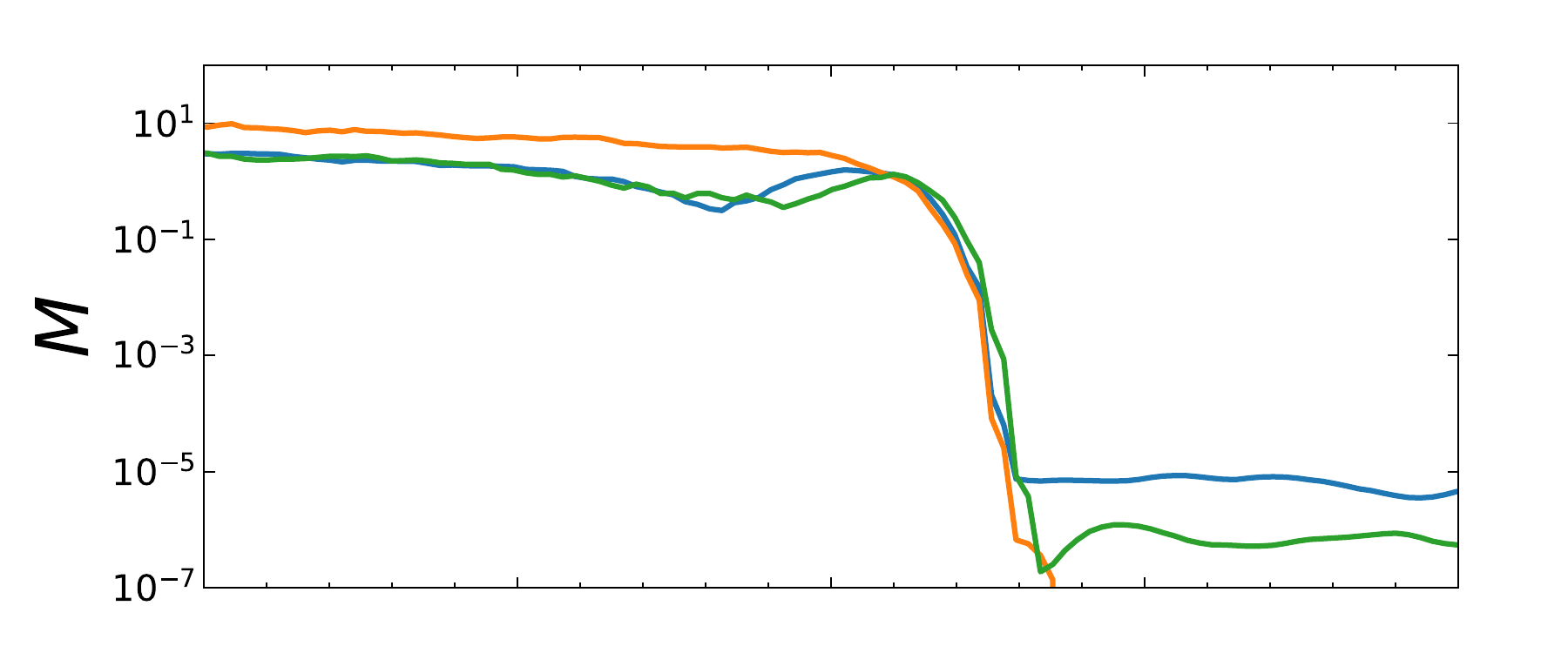}
\includegraphics[width=\columnwidth]{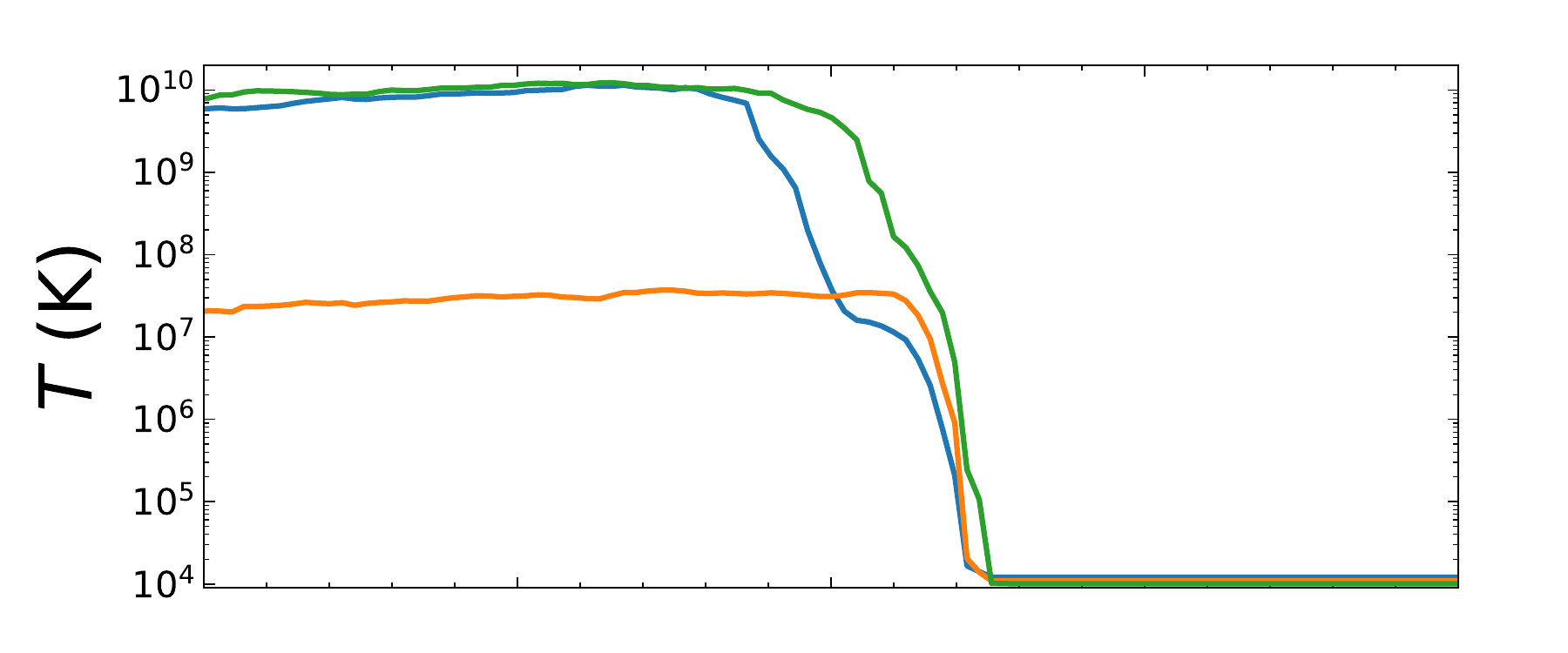}
\includegraphics[width=\columnwidth]{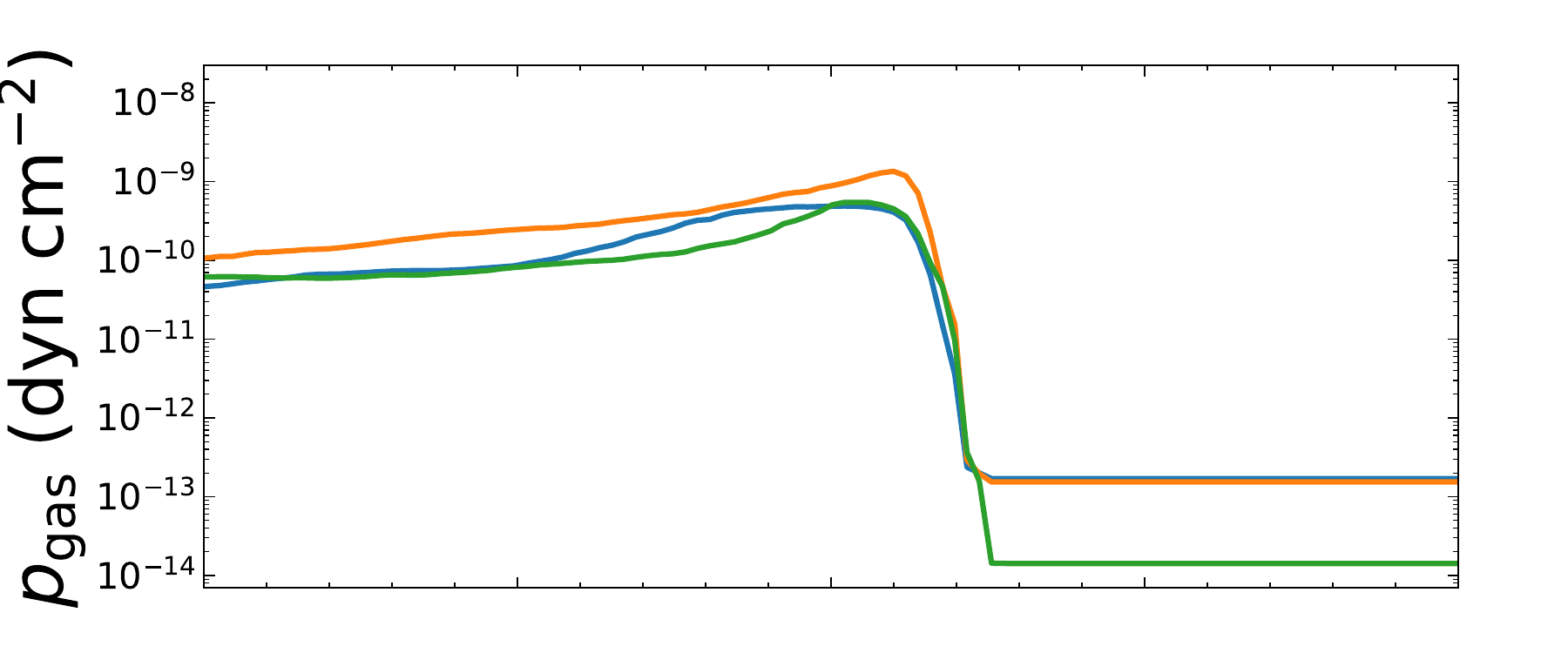}
\includegraphics[width=\columnwidth]{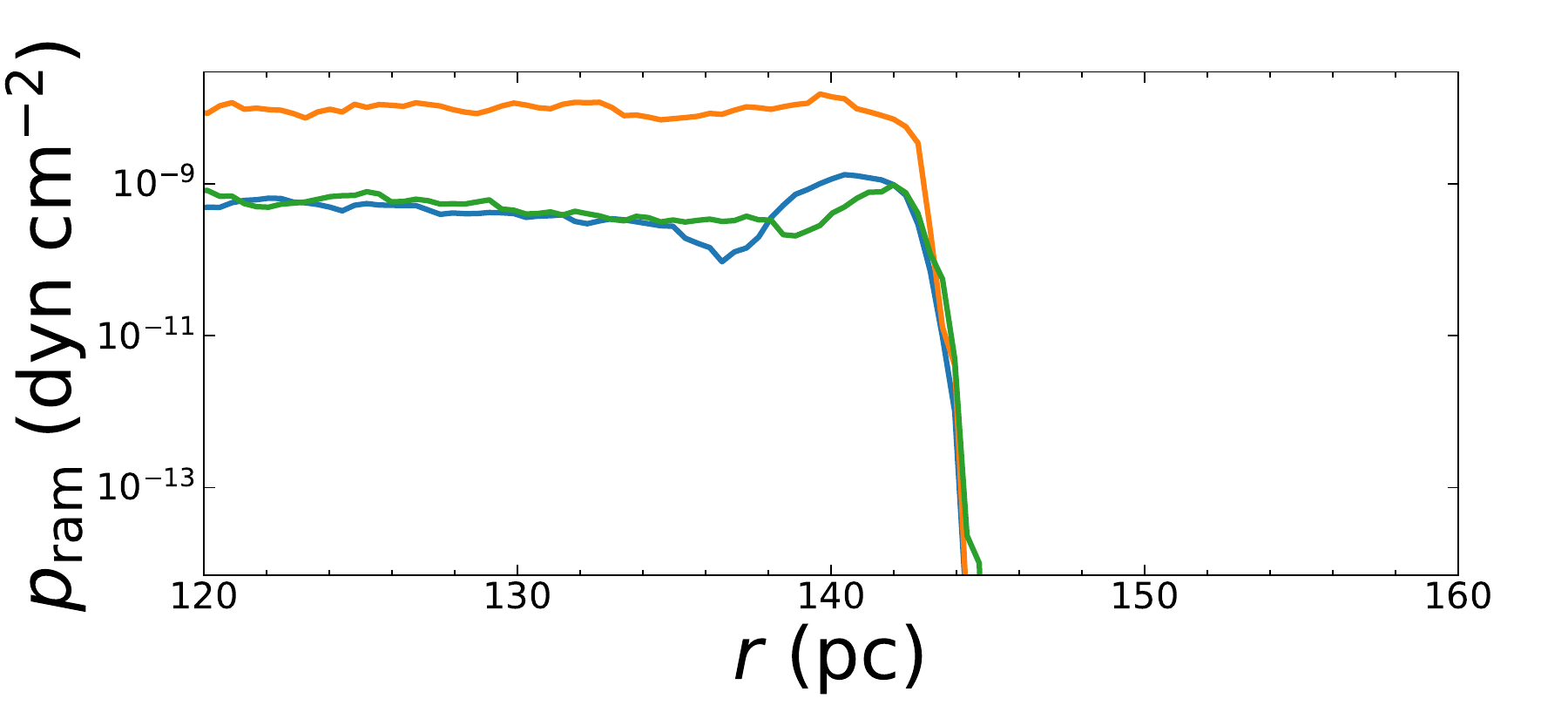}
\end{center}
\caption{Radial profiles of number density $n$, radial velocity $v_r$, local Mach number $\mathcal{M}$, gas temperature $T$, gas pressure $p_{\rm gas}$, and ram pressure $p_{\rm ram}$ of the runs JET-FID, JET-LOWV, and JET-LOWN, respectively. All the radial profiles are taken from the same snapshots of Figure~\ref{fig:jet_bubble}. The data is azimuthally and pitch angle averaged between $\theta=0^{\circ}$ and $\theta=2^{\circ}$.\\
Alt text: radial profiles of shock structures of the 4 jet runs, showing the 1D structures of number density, radial velocity, local Mach number, temperature, gas pressure, and ram pressure, respectively.
\label{fig:jet_radial}}
\end{figure}

We extract the radial profiles of the shock structures and show them in Figure~\ref{fig:jet_radial}. The radial profiles are pitch angle averaged only between $\theta=0^{\circ}$ and $\theta=2^{\circ}$ because the high velocity funnel is quite collimated.

The flow of jet inflated bubbles can also be divided into 4 parts in Figure~\ref{fig:jet_bubble}: (1) the collimated, cooler, high-velocity jet core; (2) a hot, low-density cocoon consisting of shocked jet fluid; (3) an outer, cooler, denser envelope consisting of shocked ambient gas; and (4) the undisturbed ISM. We only see a narrow peak at about 140~pc in the radial number density profile, which shows the position of the termination shock. The influence of the initial parameters upon the shock structures is still similar to the conclusions of the wind bubbles. The density and temperature of the post-shock wind are determined by the initial wind velocity and those of the termination shock front are determined by the initial ISM density. The shock fronts of all the 3 runs are propagating at a similar local Mach number.

Unlike the wind bubbles, the different phases in Figure~\ref{fig:jet_bubble} cannot be discerned in Figure~\ref{fig:jet_radial}, because the pitch angle average will make the flow phases less sharp. As a result, the shocked jet fluid and the shocked ISM near the shock front are not clearly distinguishable with a distinct contact discontinuity surface.

We see a distinct behavior of the termination shock of the run JET-LOWV. At the termination shock region, the radial velocity and gas temperature of the run JET-LOWV do not show a clear constant region, which is present in the other two runs. This is because the high momentum jet of JET-LOWV has not been fully thermalized at the termination shock and is still in the deceleration phase. This can be demonstrated if we compare the gas pressure and the ram pressure at the termination shock, where in the JET-FID and JET-LOWN runs both pressures are comparable, but in the run JET-LOWV, the ram pressure is larger than 10 times the gas pressure.

\begin{figure}
\begin{center}
\includegraphics[width=\columnwidth]{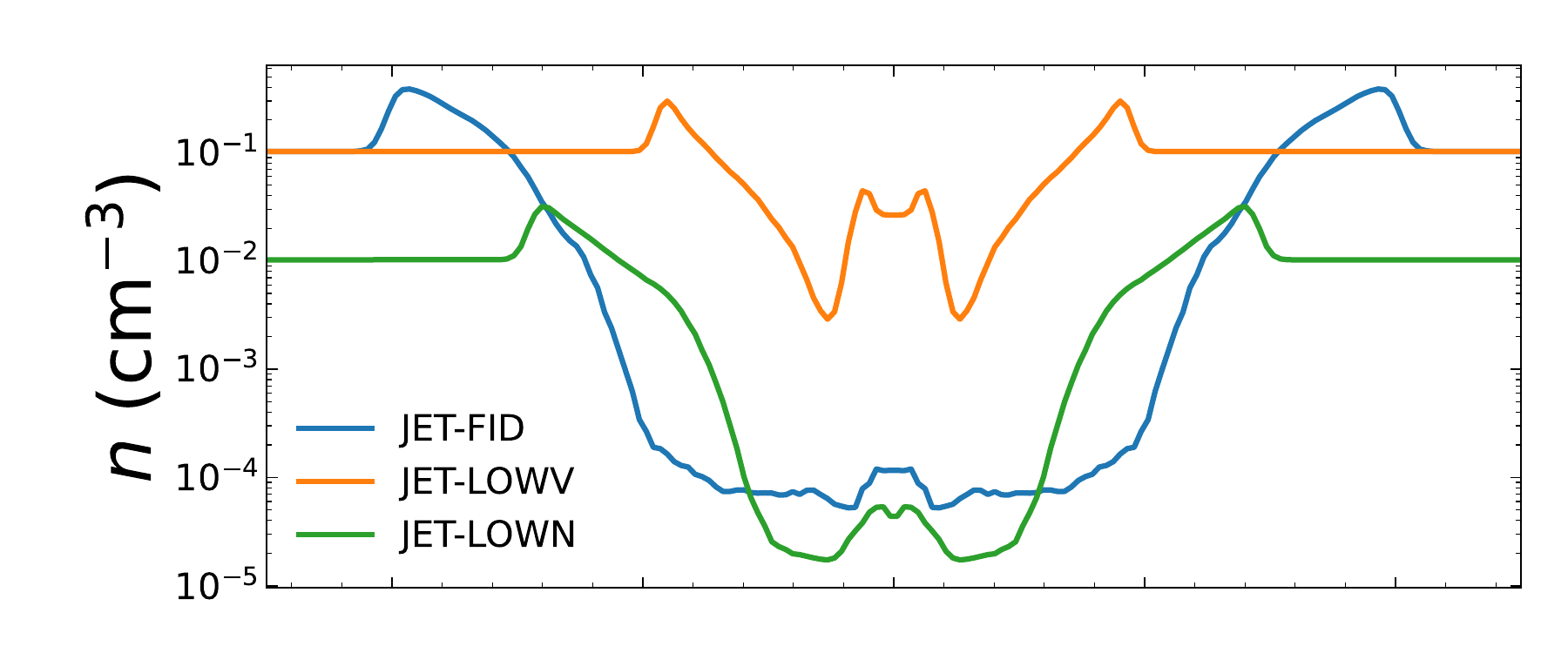}
\includegraphics[width=\columnwidth]{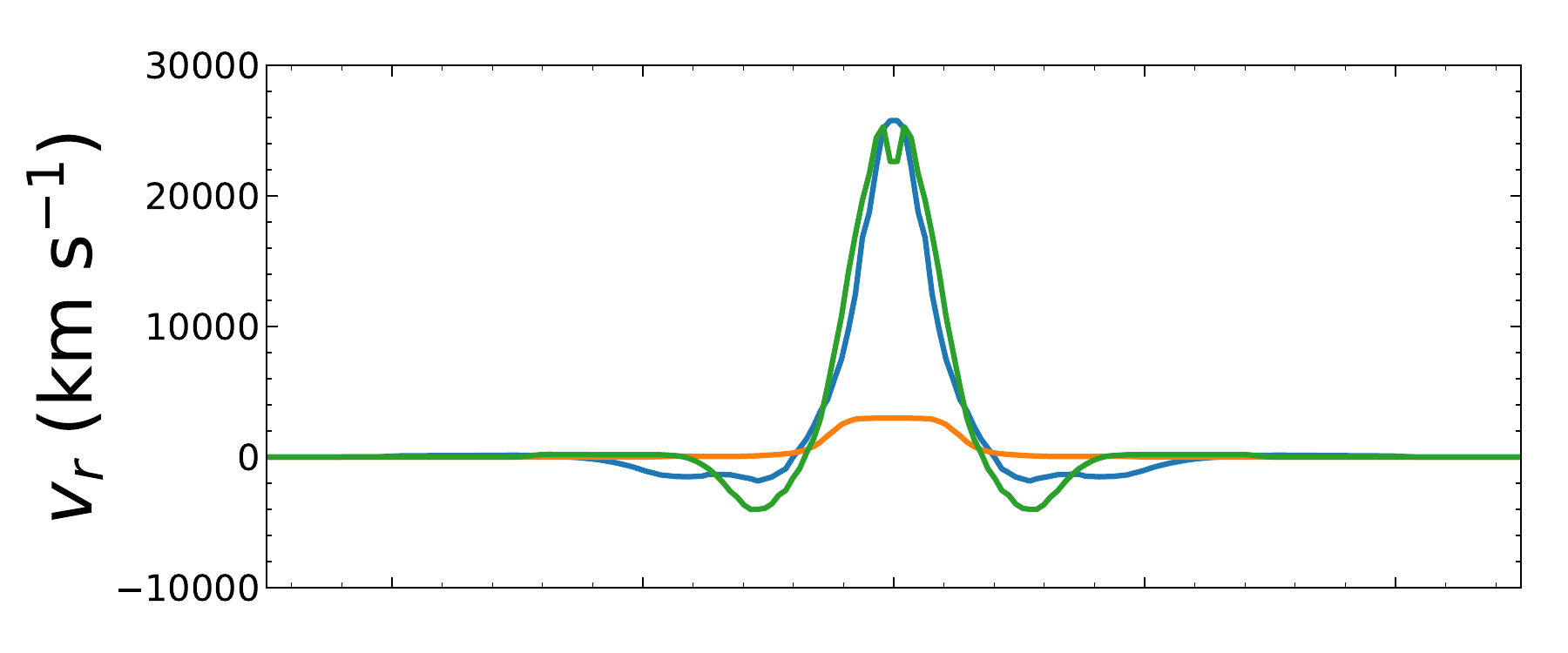}
\includegraphics[width=\columnwidth]{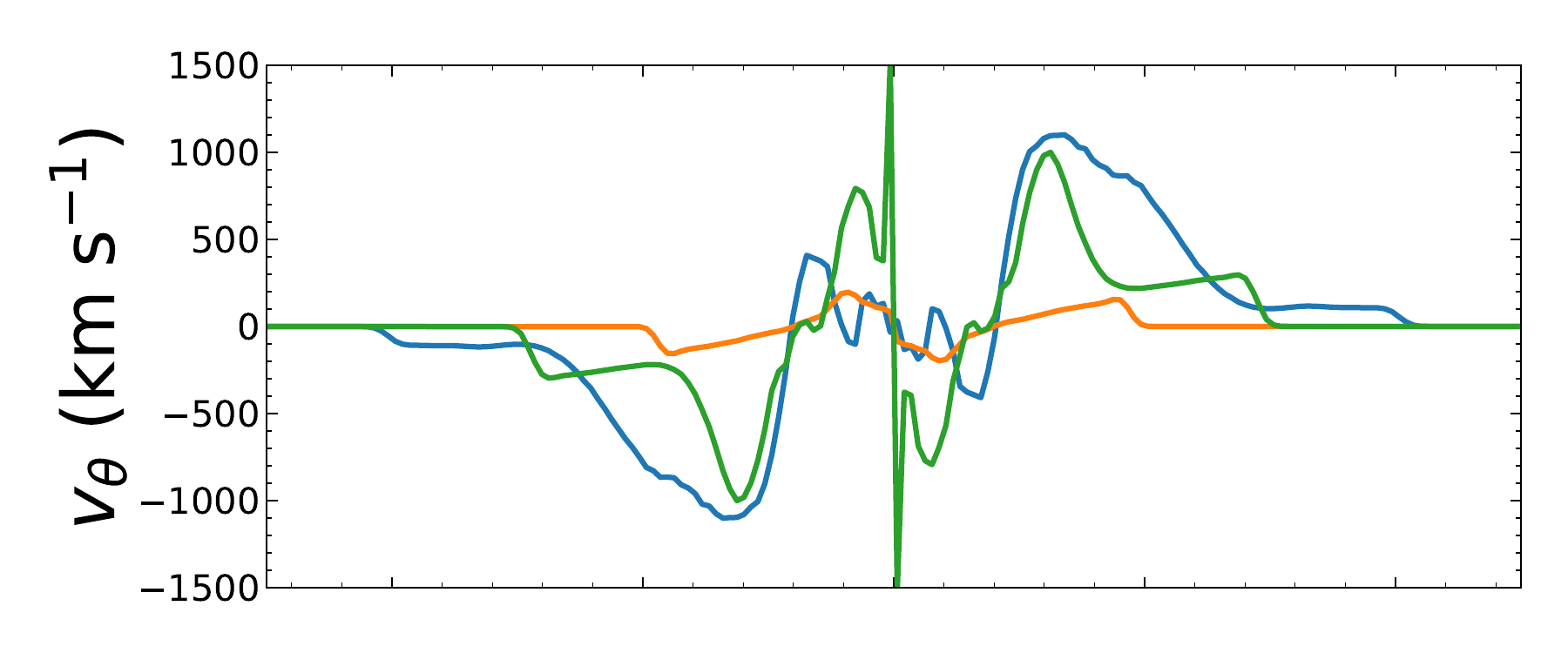}
\includegraphics[width=\columnwidth]{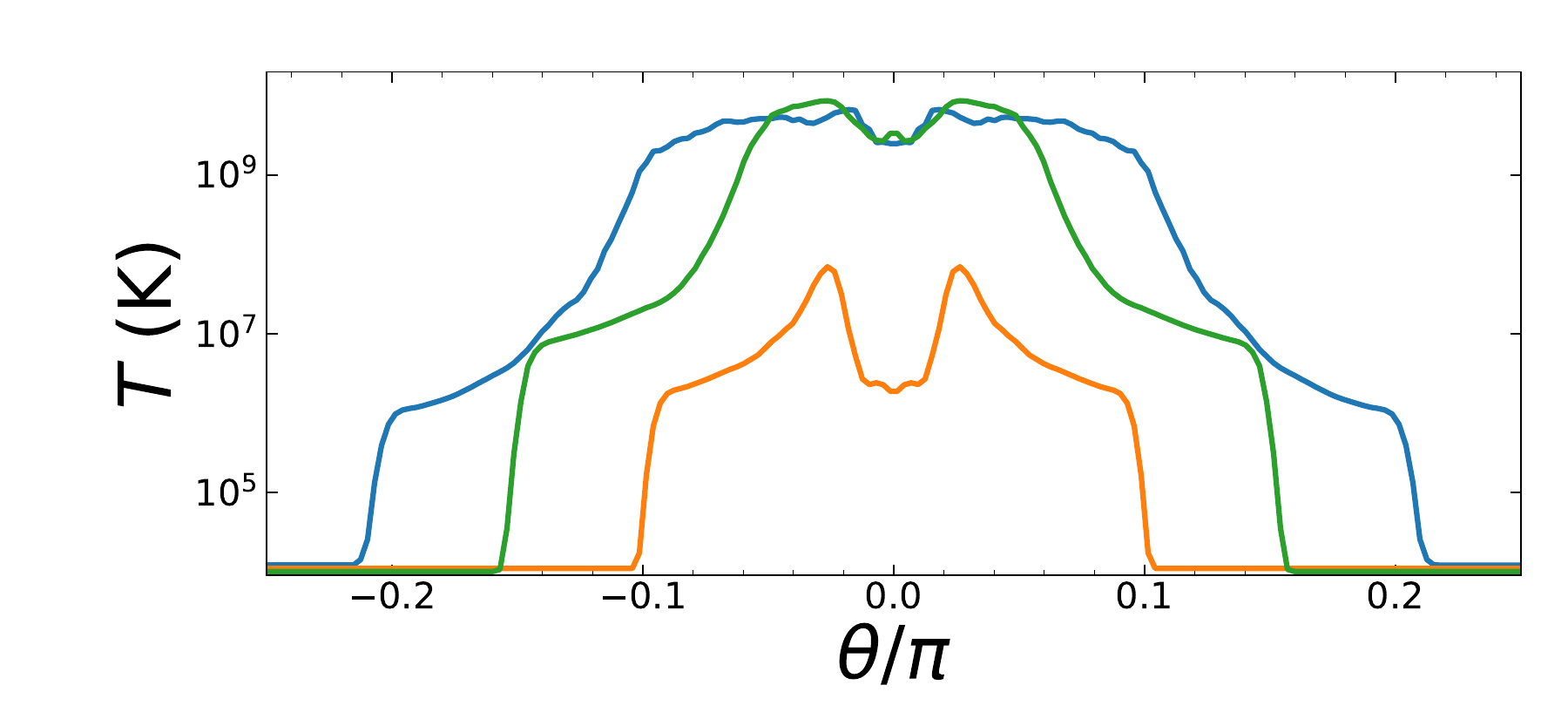}
\end{center}
\caption{Azimuthally averaged $\theta$ direction profiles of number density $n$, radial velocity $v_r$, transverse velocity $v_{\theta}$, and gas temperature $T$ of the runs JET-FID, JET-LOWV, and JET-LOWN, respectively. All the transverse profiles are taken from the same snapshots of Figure~\ref{fig:jet_bubble} and at $r=70$~pc from the central black hole.\\
Alt text: transverse profiles of bubble structures of the 4 jet runs, showing the structures of number density, radial velocity, transverse velocity, and temperature, respectively.
\label{fig:jet_transverse}}
\end{figure}

We show the transverse structures of the jet bubbles in Figure~\ref{fig:jet_transverse}, where the data is extracted 70~pc from the central accretion engine. The new feature of the jet bubbles is near $\theta=0^{\circ}$, where we see a peak in the number density profiles and a trough in the gas temperature profiles. This represents the jet cores which we have shown in the 2D plots in Figure~\ref{fig:jet_bubble}.

\section{Discussion} \label{sec:discussion}
\subsection{Influence of the outflow opening angle} \label{sec:angle}

In the third category of Table~\ref{tab:initial}, we keep the other parameters constant, changing only the half opening angle of the outflows, in order to investigate its influence on the bubble structures.

\begin{figure}
\begin{center}
\includegraphics[width=0.45\columnwidth]{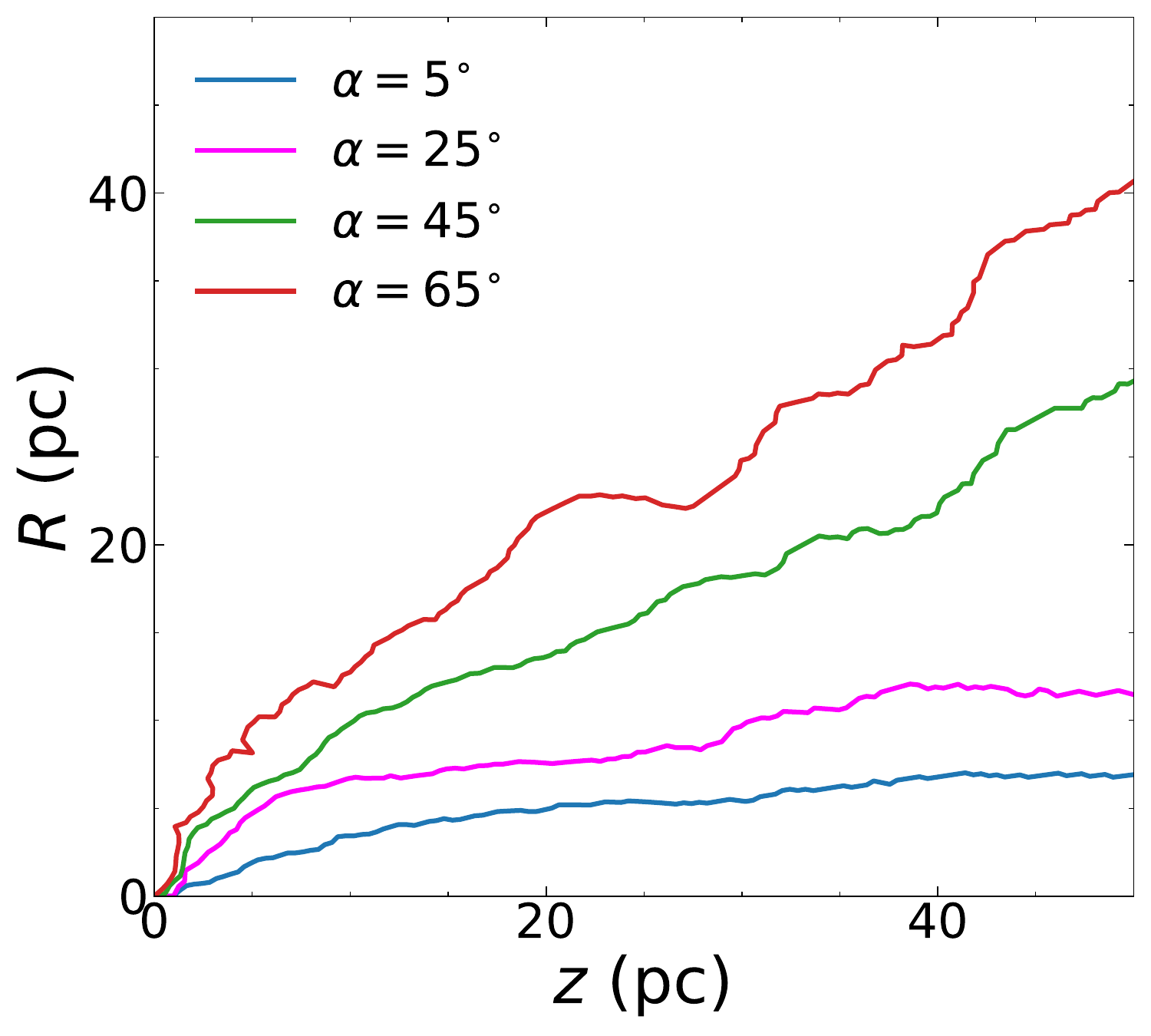}
\includegraphics[width=0.45\columnwidth]{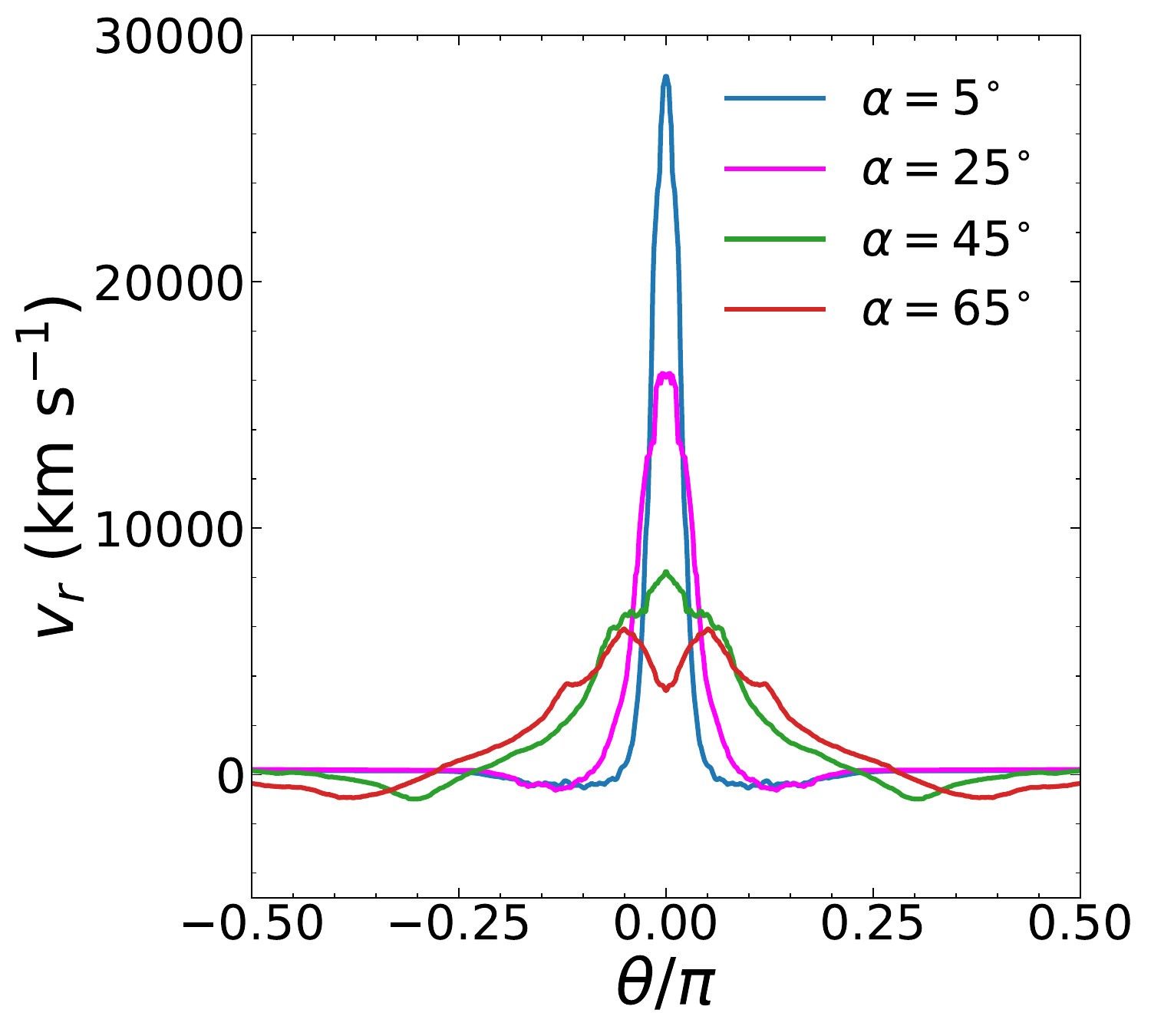}
\end{center}
\caption{{\bf Left:} Transverse radius $R$ of the high velocity region ($v_r>10^3$~km~s$^{-1}$) versus the vertical propagation distance $z$. {\bf Right:} Transverse profile of radial velocity at $z=$50~pc. The different line colors represent the results of the runs JET-FID ($\alpha=5^{\circ}$), BUB-25DEG ($\alpha=25^{\circ}$), WIND-FID ($\alpha=45^{\circ}$) and BUB-65DEG ($\alpha=65^{\circ}$), respectively.\\
Alt text: two figures showing the influence of opening angles upon the recollimation effect and the transverse propagation.
\label{fig:recollimation}}
\end{figure}

In the left panel of Figure~\ref{fig:recollimation}, we show the transverse radius $R$ of the high velocity funnel ($>1000$~km~s$^{-1}$) versus the propagation distance $z$. When the half opening angle of the outflows is large, e.g. $45^{\circ}$ or $65^{\circ}$, the radius of the funnel always increases with the propagation distance, which means that the funnel always maintains a conical shape. In contrast, for small opening angle outflows, e.g. $5^{\circ}$ and $25^{\circ}$, the funnel maintains conical before $\sim40$~pc. After that, the radius of the funnel does not change with the propagation distance, so the high velocity outflows switch from conical to cylindrical. In conclusion, the half opening angle of the outflows influences the level of recollimation. Narrow angle outflows have a stronger recollimation effect than the wide angle outflows.

We extract the transverse profiles of the radial velocity $v_r$ of the runs with different opening angles and show them in the right panel of Figure~\ref{fig:recollimation}. It is obvious that the maximum propagation speed is negatively correlated with the half opening angle, because narrow angle outflows are more collimated and more concentrated at propagating along the axis. As a consequence, the transverse profiles of the wide angle outflows are more flat, indicating more transverse propagation of the corresponding bubble structures.

\subsection{Comparison with observations} \label{sec:observation}

\begin{figure}
\begin{center}
\includegraphics[width=0.8\columnwidth]{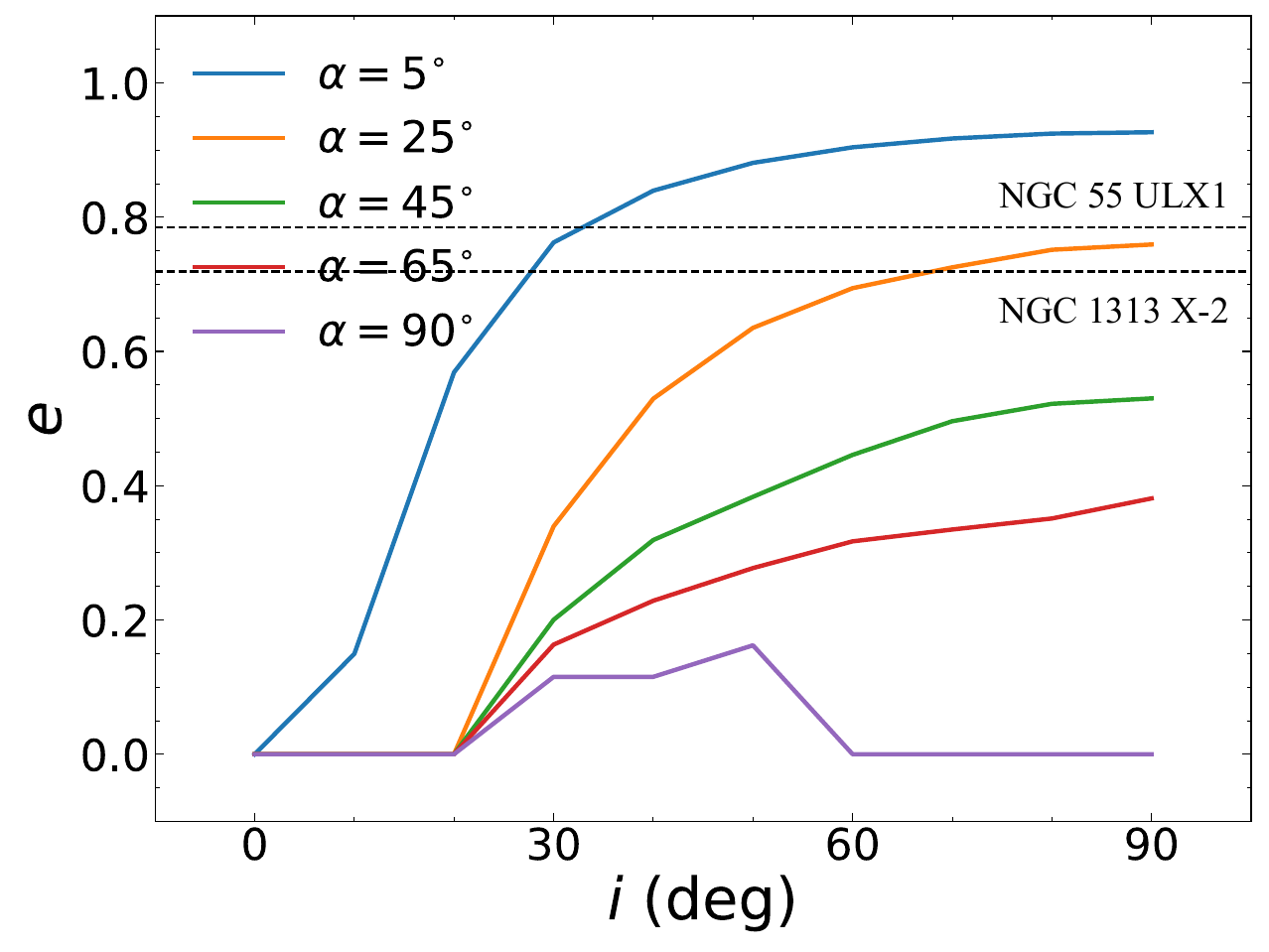}
\end{center}
\caption{Eccentricity of the bubble at different inclination angles. The different line colors represent the runs with different outflow opening angles $\alpha$. The two dashed horizontal lines show the eccentricities of the ULX bubbles NGC~55~ULX1 and NGC~1313~X-2. \\
Alt text: Joint influence of opening angles and viewing angles upon the bubble eccentricities.
\label{fig:inclination}}
\end{figure}

In ULX bubble observations, the eccentricity that observers detect is decided not only by the half opening angle of the outflows, but also the viewing angle of the system. Based on our simulation data, we place a $320\times320\times320$~pc box with different inclination angles in the simulation domains of the runs with different outflow opening angles. We calculate the emission measure of the simulated bubbles along the inclination axis of the boxes to mimic the observations with different inclination angles, where the emission measure is integrated as ${\rm EM}=\int n_{\rm e}n_{\rm p} dl$ to estimate the surface luminosity of the system. We show the viewing angle dependency of the observed bubble eccentricities in Figure~\ref{fig:inclination}. Note that in this figure, the viewing angle is defined as the angle between the line of sight and the axis of the central accretion disk, which coincides with the outflow axis.

For an ellipsoidal bubble, the observed eccentricity increases toward edge-on views with respect to the accretion-disk rotation axis, and decreases toward 0 for face-on views. When the viewing angle is fixed, the eccentricity is anti-correlated with the half opening angles of the outflows, because narrow-angle outflows create elongated bubbles whereas wide-angle outflows create circular bubbles. We mark the eccentricities of two typical elliptical ULX bubble systems in Figure~\ref{fig:inclination} with black dashed lines. For NGC 55 ULX1, two interpretations are possible: one is that the outflow has a small half opening angle ($\sim 5^{\circ}$) and the system is viewed relatively face-on ($i\sim 30^{\circ}$); the other is that the outflow has a half opening angle of about $25^{\circ}$ and the system is viewed close to edge-on. However, since this source is a supersoft source, a larger inclination angle is more likely, and thus the half opening angle of the outflow should be close to $25^{\circ}$. For NGC~1313~X-2, although we do not know the viewing angle of this system, we can at least set an upper limit of $25^{\circ}$ to the half opening angle of the outflows due to its large eccentricity. In a word, our simulation results favor that the high velocity outflows are constrained in narrow half opening angles for NGC~55~ULX1 and NGC~1313~X-2.

\subsection{Degeneracy of half opening angle and viewing angle} \label{sec:degeneracy}

\begin{figure}
\begin{center}
\includegraphics[width=0.45\columnwidth]{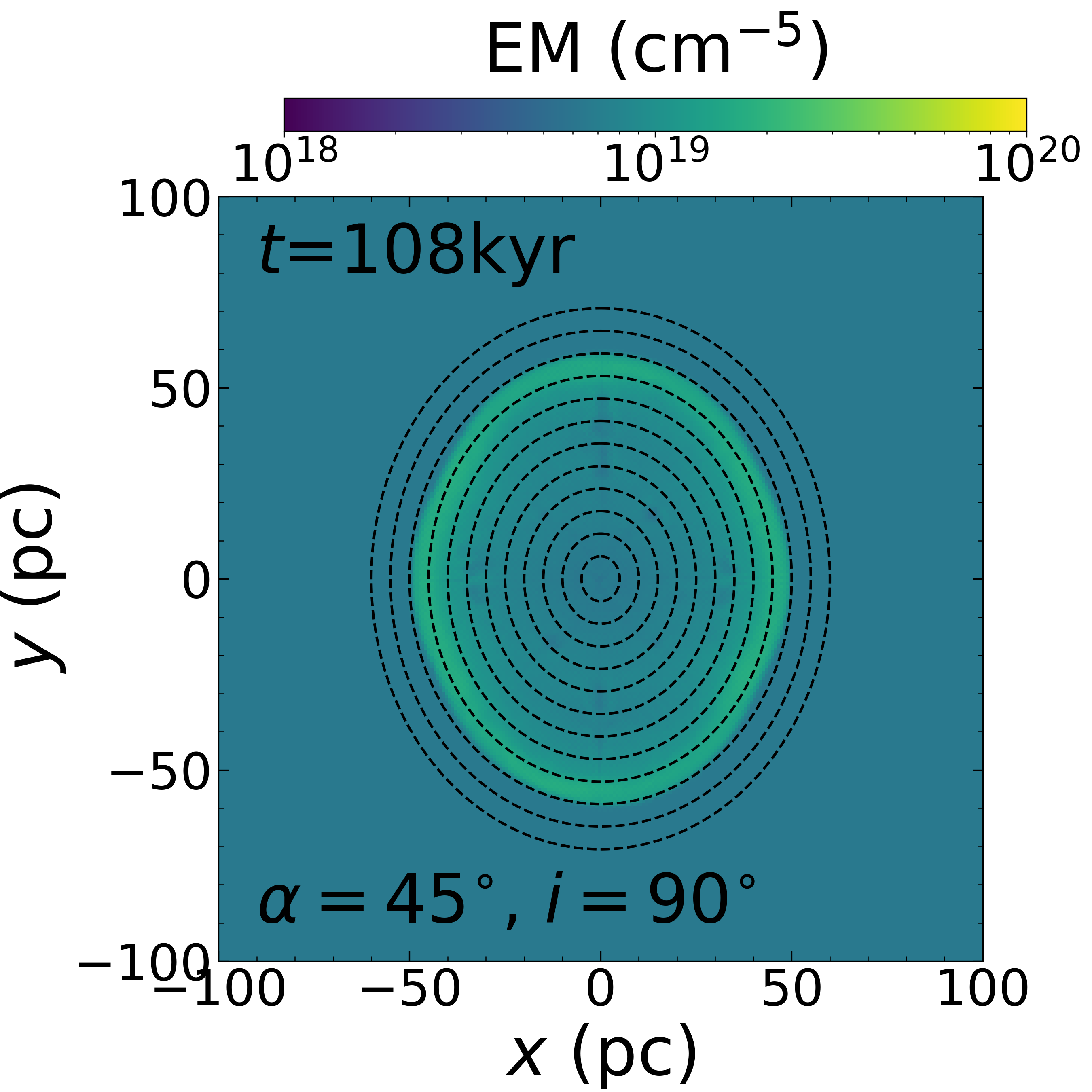}
\includegraphics[width=0.45\columnwidth]{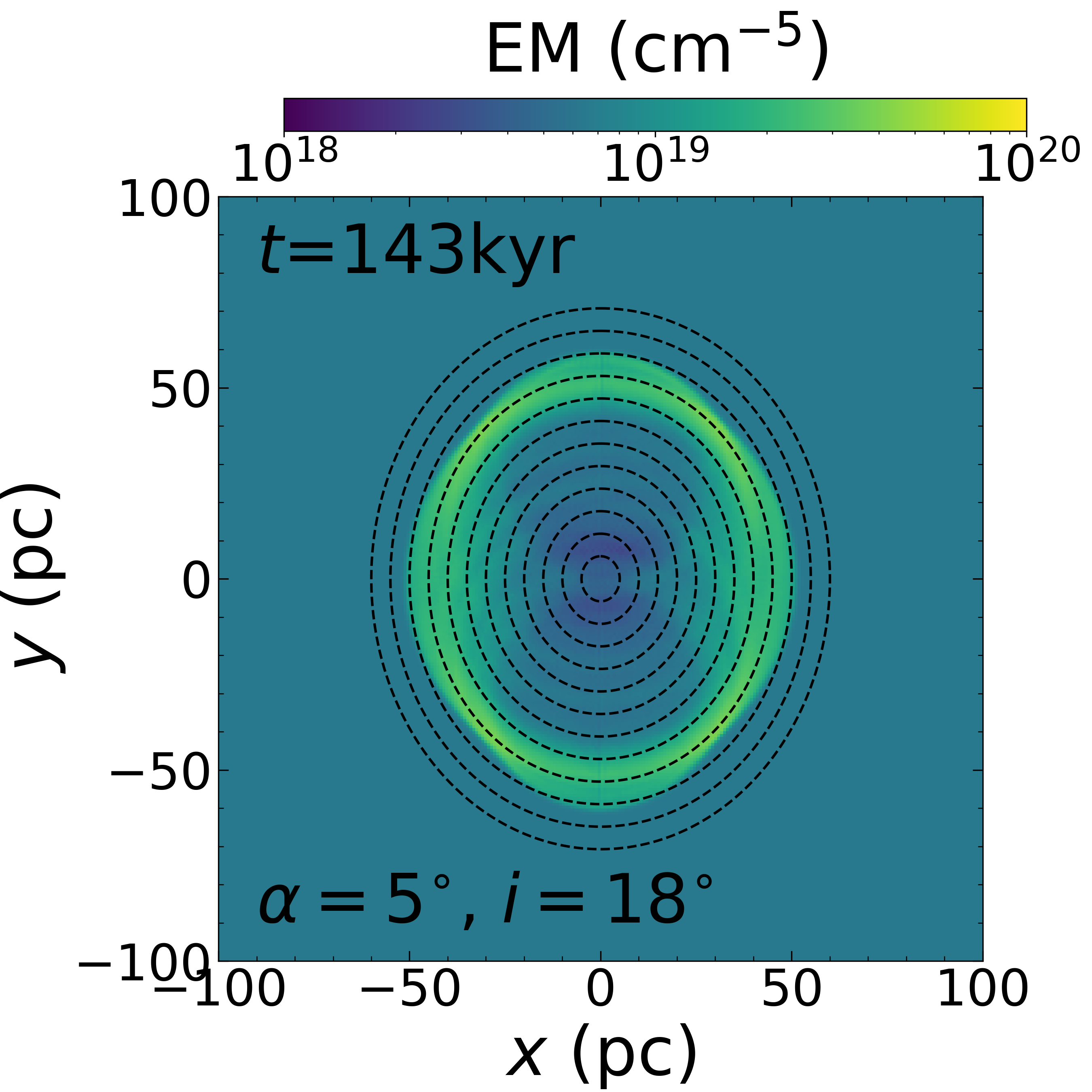}
\includegraphics[width=0.6\columnwidth]{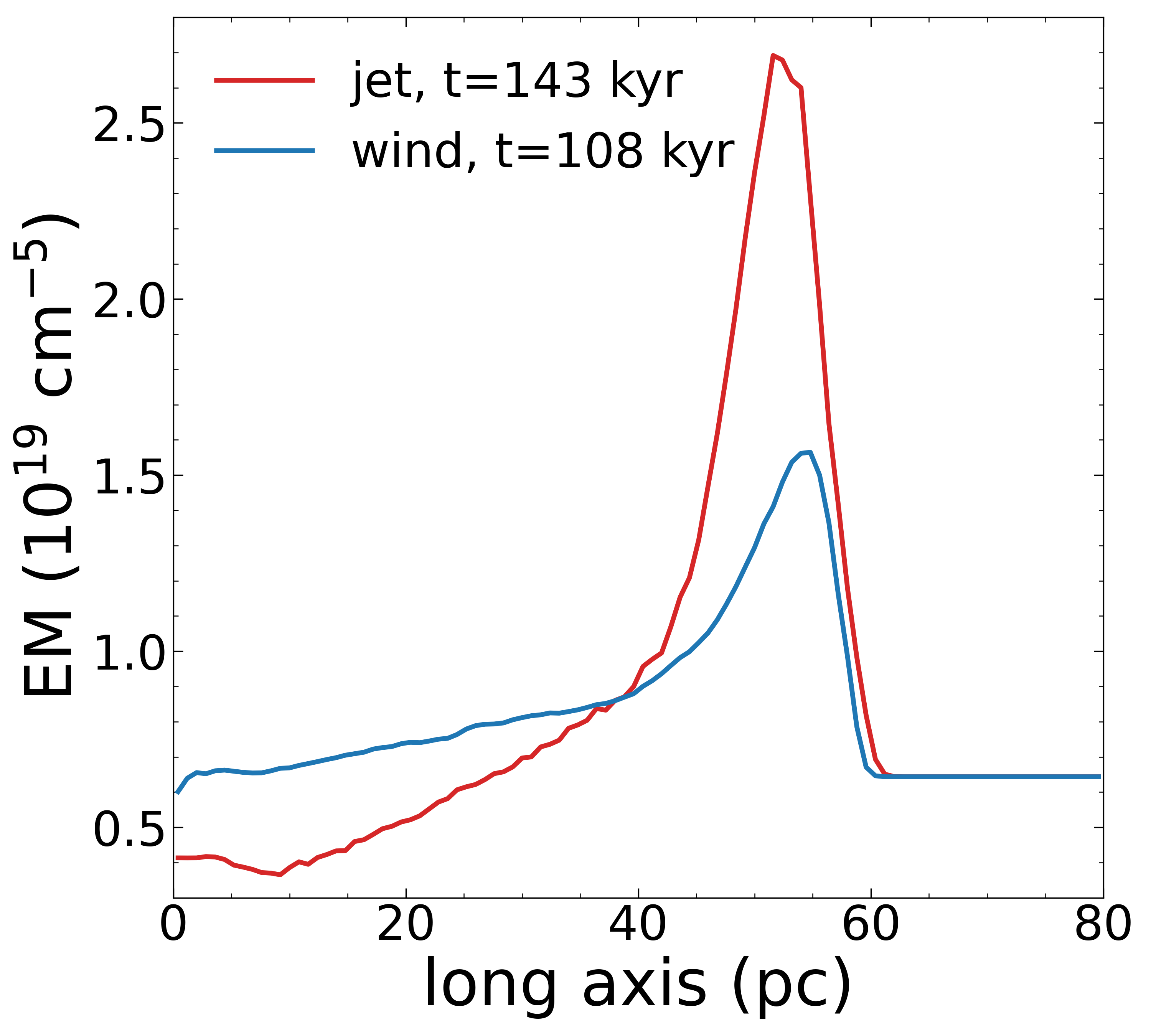}
\end{center}
\caption{{\bf Upper}: 2-dimensional emission measure maps of two bubbles with similar size and eccentricity. The left panel is a $45^{\circ}$ outflow driven bubble at a viewing angle of $90^{\circ}$. The left panel is a $5^{\circ}$ outflow driven bubble at a viewing angle of $18^{\circ}$. The dashed lines in the two panels show a series of elliptical contours with $e=0.53$. {\bf Lower}: 1-dimensional emission measure profiles of the two scenarios extracted from the contours shown as the dashed lines in the upper row. The blue and red lines represent the profiles of the jet bubble and wind bubble, respectively.\\
Alt text: degeneracy of the bubble eccentricity revealed by two different scenarios driven by wind and jet, respectively. The degeneracy can be distinguished by analyzing the 1D emission measure distributions of the two scenarios.
\label{fig:EM}}
\end{figure}

In Figure~\ref{fig:inclination}, we see a clear degeneracy between the outflow half opening angles and the viewing angles. For example, a bubble driven by $45^{\circ}$ outflows viewed at $i=90^{\circ}$ has the same eccentricity ($e=0.53$) as a bubble driven by $5^{\circ}$ outflows viewed at $i=18^{\circ}$. We select two snapshots of the runs WIND-FID and JET-FID, viewed at $90^{\circ}$ and $18^{\circ}$ respectively, so that the sizes and eccentricities of the two bubbles are similar. We show the 2D emission measure maps of the two snapshots in the upper row of Figure~\ref{fig:EM}.

Although we cannot distinguish these two scenarios by the morphology, we still see a clear difference between these two emission measure maps. The contrast between the shell and the center of the bubble is greater for the jet bubble than for the wind bubble. We further extract the 1D emission measure curves of these two scenarios using a series of elliptical contours with $e=0.53$ shown as the dashed contours in the upper row. We note here that the contours shown in Figure~\ref{fig:EM} are sparser than those we use to extract 1D emission measure curves, in order to improve visibility. The radial profiles of the emission measure are shown in the lower panel of Figure~\ref{fig:EM}. Analytically, we see that the peak value of the emission measure of the jet bubble is almost twice the peak value of the wind bubble. In addition, the emission measure inside the jet bubble shell is a bit lower than that of the wind bubble. In conclusion, we may use the combination of the observed eccentricities and the radial luminosity profiles of the ULX bubbles to distinguish the degeneracy between the outflow opening angles and the viewing angles.

\subsection{Time evolution of bubble morphology}  \label{sec:time_evolution}

One important phase of the wind driven bubble nebulae before the adiabatic evolution is the free expansion. The free expansion radius is the place where the wind density is comparable to the background ISM density. Beyond the free expansion radius, the expansion of the disk wind will be mostly thermal driven and produce rounder bubbles \citep{Schiano1985, Costa2020}. In all of our simulation runs, WIND-LOWV contains the largest free expansion radius, which is $r_{\rm free}=\sqrt{\frac{\dot{M}_{\rm w}}{4\pi\rho_{\rm ISM}v_0}} =\sqrt{\frac{2L_{\rm mec}}{4\pi v_0^3 n_{\rm ISM} \mu m_{\rm p}}}\approx1.74$~pc, comparable to the distance from the injection spheres to the black hole. As a result, our simulations cannot resolve the free expansion phase and only focus on the evolution after when the adiabatic expansion begins. However, the evolution of bubble morphologies before and after the free expansion can be quite different, e.g. the bubble may be more eccentric in the free expansion phase than that after the free expansion. We leave the simulation of small scale evolution of the accretion disk winds to a future work.

\begin{figure}
\begin{center}
\includegraphics[width=0.7\columnwidth]{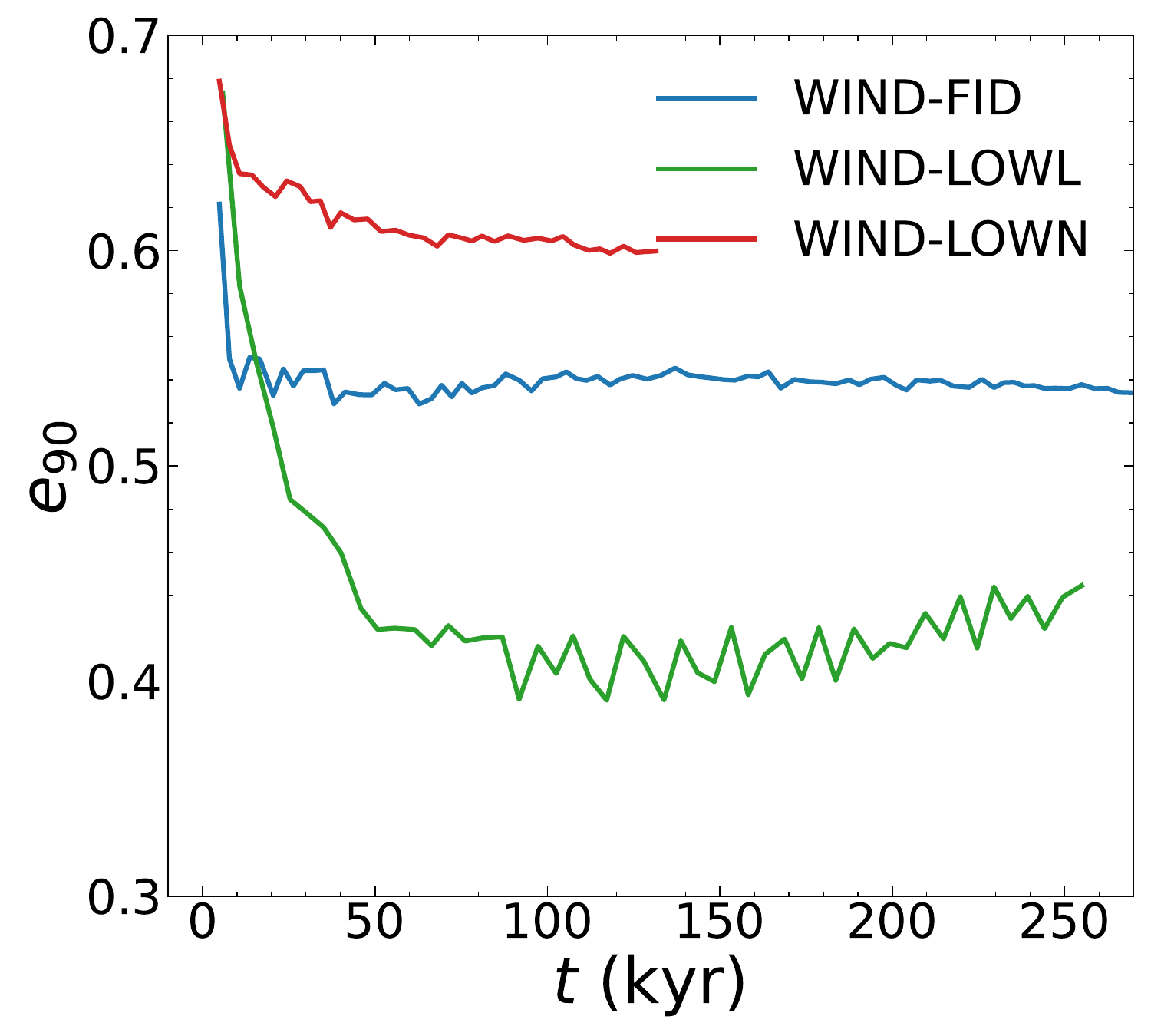}
\end{center}
\caption{Time evolution of the edge-on bubble eccentricities of the runs WIND-FID, WIND-LOWL, and WIND-LOWN.\\
Alt text: time evolution of bubble eccentricity.
\label{fig:time_evolution}}
\end{figure}

In Figure~\ref{fig:time_evolution}, we show the time evolution of the edge-on bubble eccentricities of the three wind runs. It is clear that the eccentricities have reached a relatively constant value after 50~kyr, which is because the bubble expansion has long passed the free expansion phase and becomes self-similar in the large scale. As a result, the conclusions about bubble morphologies in this paper is not time dependent. We note here that before 50~kyr, the high bubble eccentricities are because the bubble structures near the equatorial plane are not fully developed, but not because of the free expansion discussed before.

\section{Conclusion} \label{sec:conclusion}

In this work, we perform 3D moving-mesh hydrodynamical simulations of accretion disk outflow driven bubbles around ULXs, using the state-of-the-art software AREPO. Through our parameter space survey, we can summarize the characteristics of the ULX bubbles as follows.

Generally, wide angle outflows with a half opening angle of $45^{\circ}$ can produce an elliptical ULX bubble (wind bubble). However, the morphology of the bubble greatly depends on the initial momentum of the outflows. Outflows with high momentum tend to have more transverse propagation, which makes the bubble cylindrical instead of elliptical. The mechanical power of the outflows determines the size of the bubble without changing the bubble shape. Outflows with lower mechanical power produce a smaller bubble, and the smaller cooling timescale of the system leads to an early collapse of the bubble shell. The number density of the background ISM mainly influences the propagation speed of the bubble, while the temperature of the ISM has a negligible influence on the structures of the bubble before it dissipates into the ISM.

The shock structures of the simulated bubbles can be divided into the post-shock wind, the contact discontinuity, the shocked ISM, and the unperturbed ISM. The number density and temperature of the post-shock wind depend only on the initial parameters of the disk winds. Post-shock disk winds with lower initial velocity have a higher number density and lower gas temperature. The density of the shocked ISM is positively correlated with the initial number density of the background ISM, and is propagating at a similar local Mach number $\mathcal{M}\sim2$ in all simulation runs. The accretion disk winds are thermalized before reaching the shocked ISM region.

Narrow angle outflows with a half opening angle of $5^{\circ}$ can produce an elongated bubble (jet bubble). The dependency of initial parameters on the bubble properties is similar to the wide angle outflow bubbles. However, the initial momentum of the bubble influences not only the eccentricity but also the propagation speed of the termination shock front. In the jet bubbles, a low temperature region appears near the system axis and extends to about 100~pc. This is the jet core, which is not completely thermalized before it is disturbed by the turbulence.

In our simulation, we see a clear recollimation effect of the outflows due to the returning gas along the bubble shell surface. The recollimation effect is stronger if the half opening angle of the outflows is narrower. The eccentricity of the simulated bubble depends mainly on the half opening angle of the outflows and the viewing angle of the system. Generally, the eccentricity of the bubble is anti-correlated with the half opening angle of the outflows, and positively correlated with the viewing angle. Through our simulation data, it is favored that the high velocity outflows of NGC~55~ULX-1 and NGC~1313~X-2 are limited in a narrow half opening angle funnel region.

\begin{ack}
This research used resources of the CfCA Computing Facility XD2000 of National Astronomical Observatory of Japan (NAOJ). Part of the early test simulations used resources of the Orion computing cluster of Department of Astronomy, Tsinghua University, and resources of the Yukawa-21 computer system of the Yukawa Institute for Theoretical Physics (YITP). 
\end{ack}

\section*{Funding}
This work was supported by JSPS KAKENHI Grant Numbers JP21H04488, 24K00678, and 25K01045 (KO).
This work was also supported by MEXT as “Program for Promoting Researches
on the Supercomputer Fugaku” (Structure and Evolution of the Universe Unraveled by Fusion of Simulation and AI; Grant Number JPMXP1020240219; KO), 
by Joint Institute for Computational Fundamental Science (JICFuS, KO), and
(in part) by the Multidisciplinary Cooperative Research Program in CCS, University of Tsukuba (JH, KO). HF acknowledges funding support from the National Natural Science Foundation of China under the grant 12025301, and the Strategic Priority Research Program of the Chinese Academy of Sciences. HL is supported by the National Key R\&D Program of China No. 2023YFB3002502, the National Natural Science Foundation of China under No. 12373006 and 12533004, and the China Manned Space Program with grant No. CMS-CSST-2025-A10.


\bibliographystyle{astron}
\bibliography{ref}

\end{document}